\documentclass[11pt]{article}
\usepackage[toc]{appendix}

\usepackage{titlesec}
\titlespacing*{\section}{0pt}{0.7\baselineskip}{0\baselineskip}
\titlespacing*{\subsection}{0pt}{0.7\baselineskip}{0\baselineskip}
\titlespacing*{\subsubsection}{0pt}{0.7\baselineskip}{0\baselineskip}

\usepackage[utf8]{inputenc}
\usepackage{lmodern}

\usepackage{authblk}

\usepackage{xcolor}

\usepackage{amsmath}
\usepackage{amssymb} 
\usepackage{mathtools}	
\usepackage{bm}
\usepackage[makeroom]{cancel}

\usepackage[english]{babel}

\usepackage{graphicx}
	\graphicspath{{eps_fig/}}
\usepackage{subcaption}
\usepackage{setspace}  
\usepackage{epsfig}	

\usepackage{multirow}
\usepackage{booktabs} 
\usepackage{array}
\newcolumntype{C}[1]{>{\centering\let\newline\\\arraybackslash\hspace{0pt}}m{#1}} 
\newcolumntype{P}[1]{>{\raggedright\arraybackslash}p{#1}} 

\usepackage{nomencl}
\makenomenclature

\usepackage{cite}

\usepackage[colorlinks=true,
	linkcolor=blue,
	citecolor=blue,
	urlcolor=blue,
	filecolor=blue,
	pdfstartview=FitH]{hyperref}
	
\usepackage{fancyvrb} 
\usepackage{soul} 
\usepackage{enumitem}

\usepackage[bottom]{footmisc} 

\usepackage{lineno}

\definecolor{tgreen}{rgb}{0.0, 0.5, 0.0}

\begin{document}

\title{\vspace{-2.0cm}\textbf{Compliant Constant Output/ Input Force Mechanisms ---  Topology Optimization with Contact} }
\date{}
\author[$*$]{B V S Nagendra Reddy}
\author[$*$,\S]{Vitthal Manohar Khatik}
\author[$\S$]{Burkhard Corves}	
\author[$*$,$\dagger$\S]{Anupam Saxena}
\affil[$*$]{\normalsize Department of Mechanical Engineering, IIT Kanpur, 208016, India}
\affil[$\S$]{\normalsize Institute of Machine Theory, Machine Dynamics and Robotics, RWTH University, Aachen, 52074 Germany}

\affil[$\dagger$]{\small Corresponding author: \url{anupams@iitk.ac.in} }

\maketitle

\begin{abstract}
	{\it 
		{
		We synthesize monolithic topologies of constant output (CoFM) and input (CiFM) force mechanisms. During synthesis, we capture all possible aspects of member deformation including finite displacements, buckling, interaction between members, their interaction with external surfaces, and importantly, interaction of the mechanism with flexible workpieces to capture force transfer in true sense. Features of constant force characteristics, e.g., magnitude(s) of the desired force(s), range of input displacement over which slope of the force-displacement curve is near zero, and distance between workpiece and the mechanism are controlled individually via novel objectives proposed herein. Two of the constant output and constant input force mechanisms each, are synthesized using stochastic optimization ensuring ready manufacturability. We observe that presence of external surfaces may not be required for singlepiece mechanisms to attain constant force characteristics. However, interesting solutions are possible if mutual contact is permitted. We also note that desired force characteristics may not remain the same with alteration in the workpieces' shape and/or material properties. We finally fabricate and test the synthesized mechanisms and find that the desired constant force characteristics are by-and-large retained. }
	}
\end{abstract}

\section{Introduction and background}
\label{sec:introduction}

Constant Force Mechanisms (CFMs) are traditionally the ones that deliver constant \emph{output} forces while undergoing finite deformations \cite{Tolman-et-al-2016, Liu-et-al-2020}. Electrical connectors \cite{Weight-et-al-2007, Meaders-Mattson-2010}, automotive clutches \cite{Porter-1953, Li-jun-et-al-2008}, exercise equipment \cite{Wilson-1980, Howell-Magleby-2006}, robotic automation involving end-effectors/grippers \cite{Boyle-et-al-2003, Lan-et-al-2010, Chen-Lan-2012b, Wang-Lan-2014, Zhang-Xu-2019}, snap fits \cite{Chen-Lan-2012a}, MEMS \cite{Aten-et-al-2011, Wang-et-al-2011}, force regulation/overload protection \cite{Pham-Wang-2011} and precision positioning \cite{Wang-Xu-2017} are some applications they may find use in. As a CFM offers nearly constant output force (thus will be referred to as \emph{CoFM} henceforth), need for force feedback gets reduced/eliminated which makes it cost effective and easy to use, without the requirement of sensors and control systems. Some conventional (rigid-link with springs) and compliant CFMs are described in \cite{Wahl-1963, Nathan-1985, Jenuwine-Midha-1994} and \cite{Chen-et-al-2016, Zhou-Prakashah-2015, Harne-Wang-2016, Yang-Lan-2015, Wu-Lan-2015, Berselli-et-al-2009} respectively. Wang and Xu \cite{Wang-Xu-2018} provide a detailed survey, highlighting advantages and pitfalls of five types of conventional and compliant CFMs each.
Compliant mechanisms are preferred over rigid-link ones due to inherent advantages of no friction, no backlash or need for lubrication, ease of assembly and miniaturization. Thus, compliant CFMs are desired over conventional ones. Two design approaches can be employed for monolithic CFMs, just as for generic compliant mechanisms --- for instance, those in \cite{Tolman-et-al-2016} and \cite{Boyle-et-al-2003} are designed using the Pseudo-Rigid-Body Model (PRBM) approach whereas Topology Optimization (TO) is adopted in \cite{Liu-et-al-2020}.

Compliant constant input force mechanisms (\emph{CiFMs}), e.g.,  \cite{Zhang_and_Xu_2019}  can also find a variety of applications, especially if the actuation force required is \emph{ideally} zero over a range of input displacement. Design methods for, and realization of the latter class of CiFMs, termed \emph{statically balanced compliant mechanisms} (or \emph{SBCMs}), have gained significant attention in recent years \cite{Herder-1998, Tolu-et-al-2010, Hoetmer-et-al-2010, Leishman-et-al-2010, Kim-Ebenstein-2012, Stapel-Herder-2004, Cheng-et-al-2015, Lamers-et-al-2015, Merriam-Howell-2015, Deepak-et-al-2016}. Static balancing is achieved by ensuring that the (potential) energy stored within the deforming continuum is constant \cite{Hoetmer-et-al-2010} so that both, the actuation force\footnote{first derivative of potential energy w.r.t the input dof} and stiffness\footnote{first derivative of the input force w.r.t the input dof} are zero over a range of input displacement \cite{Juan-2013}. Synthesis approaches for SBCMs employ the Pseudo-Rigid-Body approach, e.g., \cite{Tolman-et-al-2016}, stiffness compensation using building blocks, e.g., \cite{Hoetmer-et-al-2010} and topology optimization, e.g, \cite{Juan-2013}. A compliant constant output force mechanism (CoFM) need not always be a constant input force mechanism (CiFM) unless no region of the mechanism is fixed so that constant forces at the two ends balance out (e.g., Fig. 2 in \cite{Tolman-et-al-2016}). Two such CiFMs, as building blocks, can be pre-loaded, mirrored and fastened at their actuation ends resulting in an SBCM with virtually no opposition to motion at the combined input, as demonstrated in \cite{Tolman-et-al-2016}. Literature, so far, has not formally distinguished between compliant constant input and output force mechanisms. In this paper, such a distinction is maintained and CFMs are specifically referred to as either CoFMs or CiFMs. CoFMs and CiFMs can be referenced with respect to either input or output displacements. We employ input displacement as reference (see Fig.  \ref{fig:objective_main}) as output displacements would usually vary with the type and material properties of the workpiece.

Few methods to synthesize CoFMs and CiFMs based on continuum optimization exist \cite{Pedersen_Fleck_and_Suresh_2006, Juan-2013, Liu-et-al-2020,Liu_Chung_Ho_2021}. Pedersen et al. \cite{Pedersen_Fleck_and_Suresh_2006} perform topology optimization using 2D continuum parameterization and size optimization on frame element abstraction of the corresponding solution, to design CoFM for given linearly decreasing force vs displacement characteristic.  In \cite{Juan-2013}, topology optimization is performed with frame elements using stochastic search, with two different optimization formulations, namely, (a) neutral stability (using the buckling modes) and (b) continuous equilibrium (using large deformation analysis). Further, three different concepts are used to demonstrate a statically balanced compliant suspension (SBCS). A combination of two or more constant force mechanisms is pre-stressed to achieve the SBCS. In \cite{Liu-et-al-2020} and \cite{Liu_Chung_Ho_2021} that synthesize CoFMs, geometric and material nonlinearities are considered with rectangular cells/finite elements. To resolve mesh distortion and non-convergence related problems pertaining to low stiffness cells in large deformation finite element analysis, hyperelastic elements are added. Topology optimization is performed within the Matlab and ANSYS environments using gradient search followed by geometry optimization. In \cite{Liu-et-al-2020} and \cite{Liu_Chung_Ho_2021}, the output force is simulated via a virtual,  linear, output spring.

\section{Aim, novelty, scope and organization}
\label{sec:aim_sb_cf_cms}
We aim at designing constant output force (CoFM) and input force (CiFM) mechanisms using topology optimization such that obtained solutions are directly manufacturable \emph{as is}. Using the procedure delineated in \cite{Reddy-saxena-2020} with initially curved frame elements and zero order search, we intend to capture all geometrically nonlinear deformation modes related to (a) member buckling, (b) interaction between members (self contact), (c) interaction between members and external surfaces of specific shapes (mutual contact), (d) interaction of the continuum with workpiece, all of which will contribute, non-intuitively, towards the manner in which energy is stored within the deforming monolithic continuum. Such features, implemented herein, are not pre-specified and are determined systematically by the developed algorithm such that energy storage within the continuum can be manipulated via contact. The aforementioned  has not been captured yet in topology design of large displacement compliant, constant input/output force mechanisms. We ensure force transfer by modeling contact forces accurately between the continuum and workpiece (assumed flexible having nonlinear material properties) rather than simulating those by means of a virtual output spring (e.g., \cite{Liu-et-al-2020},  \cite{Liu_Chung_Ho_2021}). We propose and employ novel \emph{function generation} based objectives to synthesize CoFMs and CiFMs providing individual control on different design features, namely, maximized input displacement range over which the output or input forces are constants, and intended magnitudes of such forces.

\section{Problem Formulation}
\label{sec:prob_form}

\begin{figure*}[h!]
	\centering
	\begin{subfigure}[t]{0.6 \textwidth}
		\centering
		\includegraphics[scale=0.7]{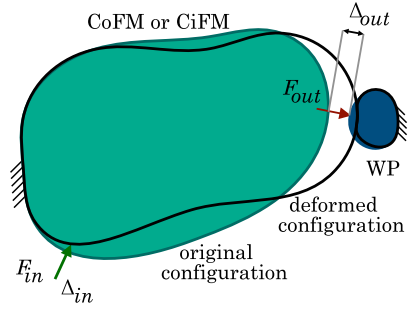}
		\caption{A candidate CiFM or CoFM continuum interacting with workpiece WP} \label{fig:objective_a}
	\end{subfigure}	\hfill \\
	\begin{subfigure}[t]{0.45 \textwidth}
\centering		\includegraphics[scale=0.3]{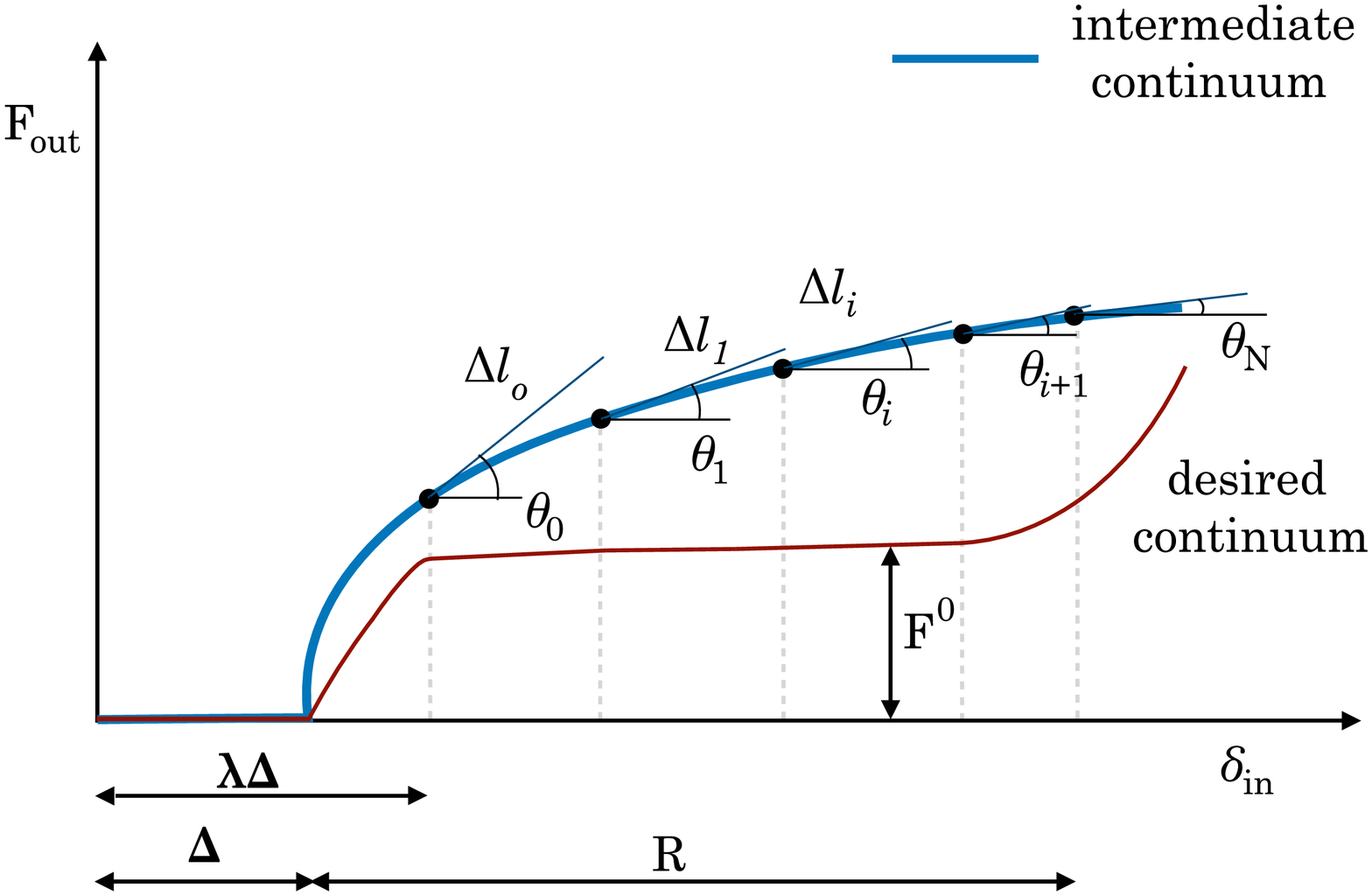}
		\caption{${F_{out}} -{\delta_{in}}$ function behavior of a candidate CoFM continuum. $ {\Delta } $: $ \Delta_{in} $, the input displacement for which the output displacement is $ \Delta_{out} $. For $\delta_{in} < \Delta_{in}$, the output force is zero. }
		\label{fig:objective_b}
	\end{subfigure} \hfill 
	\begin{subfigure}[t]{0.45 \textwidth}
\centering		\includegraphics[scale=0.3]{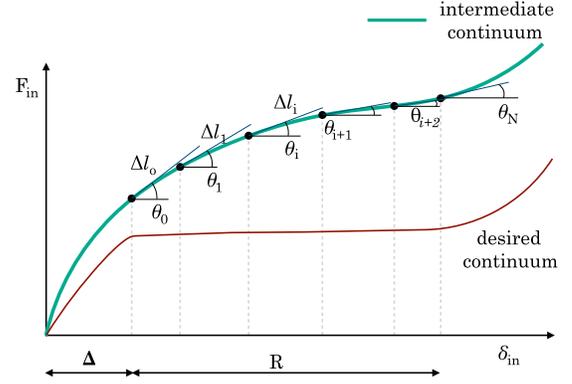}
		\caption{${F_{in}} -{\delta_{in}}$ function behavior of a candidate CiFM continuum. $ { \Delta } $: $ \Delta_{in} $, i.e.,  the displacement of input port for which the output displacement is $ \Delta_{out} $.}
		\label{fig:objective_c}
	\end{subfigure}		
	\caption{(\subref{fig:objective_a}) The undeformed and deformed configurations of both continuum and workpiece. $ F_{in} $ and $ F_{out} $ represent the actuation and output forces at the input and output ports respectively.  $\Delta_{in}$ represents the initial deformation observed at the input port, at the start of interaction between the continuum and workpiece. (\subref{fig:objective_b}-\subref{fig:objective_c}) Piece-wise segmentation of ${F} -{\delta_{in}}$ curve. $F^0$ is the respective desired constant value within the displacement range $R$.}
	\label{fig:objective_main}
\end{figure*}

\noindent Consider a compliant continuum in Fig. \ref{fig:objective_a}, in both, deformed and original configurations. Upon the application of load $ F_{in} $, the continuum deforms and comes in contact with the workpiece placed at a distance (minimum) $ \Delta_{out} $ from the output port transferring force $F_{out}$. This force, in general, is nonlinear and depends primarily on the interaction between the continuum and the workpiece, given their material properties. The input port traverses a distance $\Delta_{in}$ just before the aforementioned contact occurs. One considers the discrete piece-wise form of the ${F} -{\delta_{in}}$ (${\delta_{in}}$ is the input displacement) curve over a displacement range $ R $ and computes the slope $ \theta_i, i=0,1,2, ..., N $ of each segment. 

\subsection{Constant output Force Mechanism (CoFM)}
\label{sec:CoFM_pf}

Consider discretized form of the function response $ F_{out} - \delta_{in} $ of a candidate CoFM (Fig. \ref{fig:objective_b}). For $\delta_{in} < \Delta_{in}, F_{out} = 0$, and as $\delta_{in}$ increases, the reaction force registers a non-zero value.  For $\delta_{in} > \lambda_{in}\Delta_{in}$, where $\lambda_{in} > 1$ is some (prespecified) factor, the CoFM is required to maintain a constant, desired output (contact) force $F^{0}$ and hene near zero slope, $\sum\theta_i^2$, over a range $R$ of the input displacement. As $\Delta_{in}$ is the input displacement for an intermediate continuum for its output port to be displaced by $\Delta_{out}$, it will not be the same for all intermediate solutions. The range $R$ may be computed as the sum of lengths of piece-wise segments, i.e., $\sum{\Delta l_i}$. This range may be desired to be as large as possible. One could therefore seek an optimal topology of the compliant CoFM via minimizing the following objective:
\begin{align}
	\label{ratio_CoFM_obj}
	O_{1R}:   \displaystyle\text{min}\;\; & \sqrt{ \left[ (F_{out}-F^0)^2 \right]_{\delta_{in} = \delta_{in}^{*} } } \times \nonumber \\  
	& \left\{\frac{ \sqrt{   {\left[\sum\theta_i^2\right]}  }   }{\sum{\Delta l_i}}\right\}_{\delta_{in}>\lambda_{in}\Delta_{in}}
\end{align}

\noindent One notes that $F_{out}$ in Eq. \ref{ratio_CoFM_obj} is the reaction force obtained from the $F_{out}-\delta_{in}$ response, at a point $\delta_{in}^{*} $ just after $\delta_{in}$ becomes larger than $\lambda_{in}\Delta_{in}$. Suitable changes could be made if certain sub-objectives are to be revised or dropped. For instance, if $F_{out}$ is to be maximised, $[F_{out}-F^0]$	in $O_{1R}$ may be replaced by $\frac{1}{F_{out}}$. If having a large range is not deemed necessary, ${\sum{\Delta l_i}}$ may be dropped. Such individual control is also possible with the linear combination objective type:

\begin{align}
	\label{lc_CoFM_obj}
	O_{1L}: \displaystyle\text{min    }  \;\;\;  & \left[ C_1 \sqrt{ \sum \theta_i^2 } + C_2 \frac{1}{\sum{\Delta l_i}}\right]_{\delta_{in}>\lambda_{in}\Delta_{in}} + \nonumber \\
	& C_3  \sqrt{ \left(F_{out}-F^0\right)^2_{\delta_{in}= \delta_{in}^{*}}}
\end{align}

\begin{figure*}[t!]
	\centering
	\includegraphics[scale=0.8]{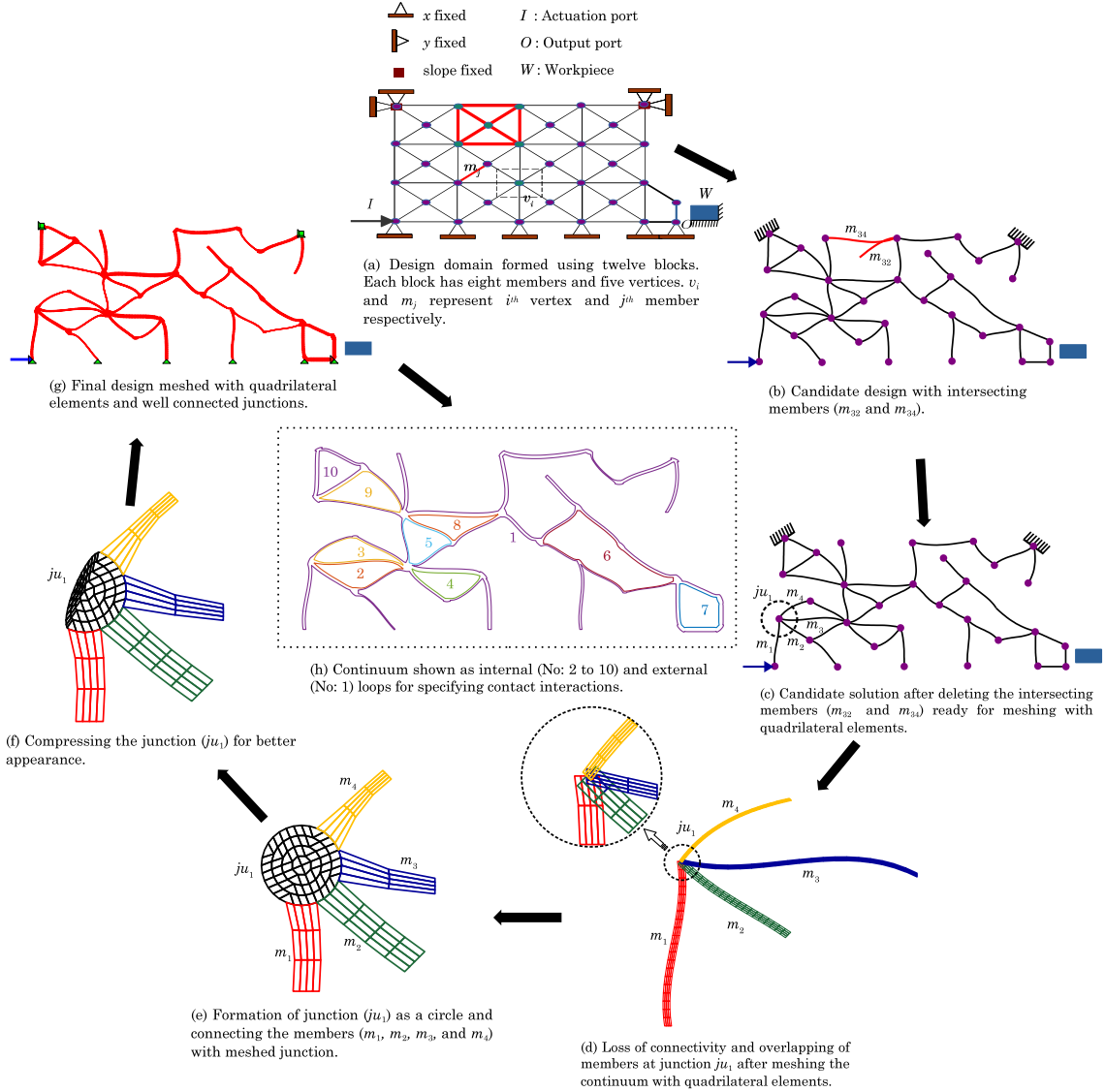}
	\caption{Processes involved in preparation of a candidate design for large deformation analysis.}
	\label{fig:DesignDomain}
\end{figure*}

\noindent	where $ C_1, C_2, \text{and }  C_3 $ are suitably chosen positive weights corresponding to the error in slopes of all segments, range ($R$) and deviation in the output force from that desired. In objective $O_{1R}$, weight specification is not required. However, $O_{1R}$ can assume a low value very quickly (irrespective of the values of $\sum \theta_i^2$ and $\frac{1}{\sum{\Delta l_i}}$) in the optimization process if $F_{out}$ gets close to $F^0$, at $\delta_{in} = \delta_{in}^{*}$, a single point on the $F_{out} - \delta_{in}$ curve. Both objective types are expected to have non-convex design spaces and can yield multiple (topological) solutions. 

We prefer to employ $O_{1L}$ over $O_{1R}$,  and the objective proposed in \cite{Liu-et-al-2020,Liu_Chung_Ho_2021}.  Therein, the workpiece is modeled with constant stiffness and $F_{out}$ computed at a single node. We model the workpiece with known geometry and hyper-elastic (nonlinear and hence non-constant output spring) material properties. In addition, we model the contact force over a surface, consider the initial gap ($\Delta_{out}$) between the output port and the workpiece, and also maximize the range of input displacement over which the output force is constant. These nonlinearities are not considered in \cite{Liu-et-al-2020,Liu_Chung_Ho_2021}.

\subsection{Constant input Force Mechanism (CiFM)}
\label{sec:CiFM_pf}	

Considering the $ F_{in} - \delta_{in} $ function curve (Fig. \ref{fig:objective_c}), the base intent behind the monolithic constant input force mechanism (CiFM) is similar to that for the CoFM. That is, slope of the $F_{in}-\delta_{in}$ curve should be close to zero in some range ($\delta_{in} > \Delta_{in}$) of the input displacement, and also that range, $R = \sum \Delta l_i$, should be as large as possible. In addition, if a CiFM is sought to possess characteristics similar to a statically balanced compliant mechanism or one of its building units, one may intend to minimize the total magnitude of input force $F_{in}$ beyond $\delta_{in} > \Delta_{in}$ ($\lambda_{in} = 1$) so that the latter is close to the horizontal axis over a range of input displacement. One may also consider maximizing the force transfer, that is, maximizing the final interactive force at the output port. In that regard, the ratio and linear combination type objectives can be formulated respectively as
\begin{align}
	\label{ratio_CiFM_obj}
	O_{2R}: & \;\;\; \displaystyle\text{min }  \;\;\; \sqrt{\left[ \frac{F_{in}^2}{F_{out}^2} \right]_{F_{out} > 0}}   \left\{\frac{{\sqrt{\sum\theta_i^2}}}{\sum{\Delta l_i}}\right\}_{\delta_{in}>\Delta_{in}} \\
\text{and} &  \nonumber \\
	\label{lc_CiFM_obj}
	O_{2L}: & \;\;\; \displaystyle\text{min    }  \;\;\; \left[ C_1 \sqrt{ \sum \theta_i^2} + C_2 \frac{1}{\sum{\Delta l_i}}\right]_{\delta_{in}>\Delta_{in}} + \nonumber \\
	&  C_3 \sqrt{F_{in}^{2}} + C_4 \left(\sqrt{\frac{1}{F_{out}^2}}\right)_{F_{out} > 0} 
\end{align}

\section{Topology Optimization}
\label{sec:to_cifm_cofm}
\begin{figure*}[t!]
	\centering
	\begin{subfigure}[h]{0.35 \textwidth}
		\centering
		\includegraphics[scale=0.6]{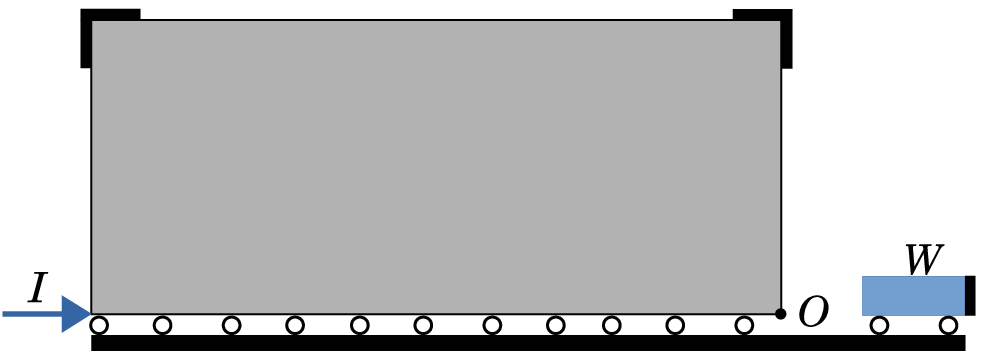}
		\caption{} 
		\label{fig:CoFM_example1_a}
	\end{subfigure}  \hfill
	\begin{subfigure}[h]{0.3 \textwidth}
		\centering
		\includegraphics[scale=0.6]{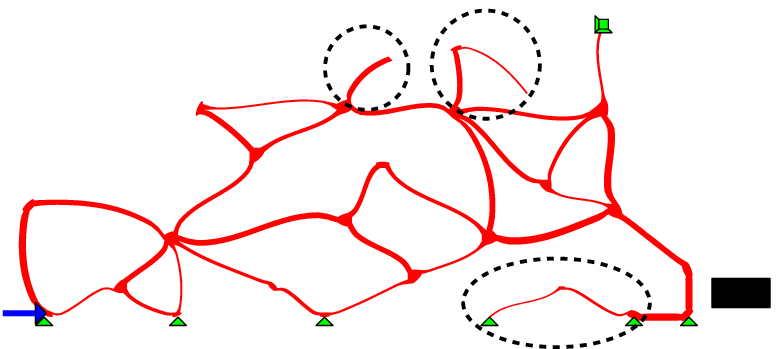}
		\caption{}
		\label{fig:CoFM_example1_b}
	\end{subfigure}    \hfill 
	\begin{subfigure}[h]{0.3 \textwidth}
		\centering
		\includegraphics[scale=0.6]{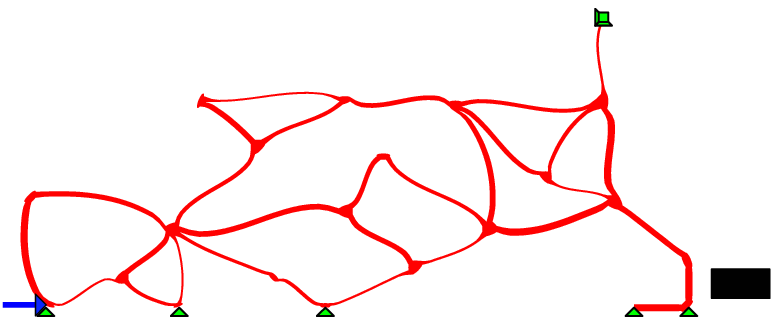}
		\caption{}
		\label{fig:CoFM_example1_c}
	\end{subfigure}   \\ 
	\begin{subfigure}[t]{0.45 \textwidth}
		\centering
		\includegraphics[scale=1]{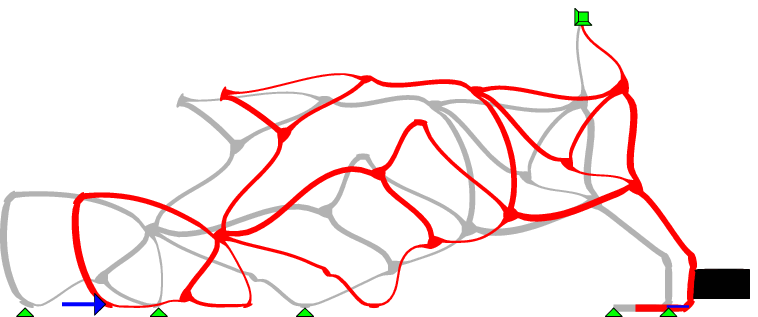}
		\caption{}
		\label{fig:CoFM_example1_g}
	\end{subfigure} 
	\begin{subfigure}[t]{0.45 \textwidth}
		\centering
		\includegraphics[scale=0.75]{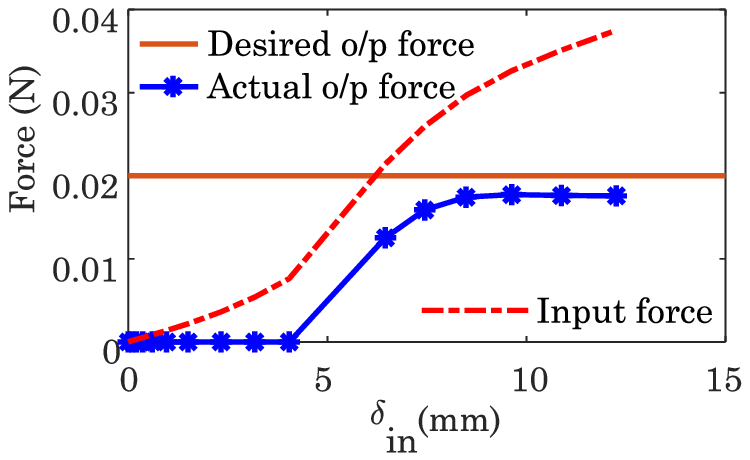}
		\caption{}
		\label{fig:CoFM_example1_h}
	\end{subfigure} 
	\caption{(\subref{fig:CoFM_example1_a}) Design domain for CoFM-I with workpiece (W). $ I: $ input port; $ O: $ output port. (\subref{fig:CoFM_example1_b}) optimized solution after 17407 function evaluations (see Fig. \ref{fig:Convergance_history}) with dangling members encircled with dashed lines. (\subref{fig:CoFM_example1_c}) Final design without dangling members. $\Delta_{\textmd{out}} = $ 4.8mm. (\subref{fig:CoFM_example1_g}) Final deformed configuration of designed CoFM (gray represents undeformed). (\subref{fig:CoFM_example1_h}) ${F_{out}} -{\delta_{in}}$ behavior of CoFM-I along with desired force.}
	\label{fig:CoFM_example1}
\end{figure*}

We summarise the methodology to create candidate solutions to prepare them for the finite element analysis. We start with a grid like structure formed by combining different blocks of straight lines (Fig. \ref{fig:DesignDomain}a). Each block comprises of \textit{eight} members ($ m_j $) and \textit{five} vertices ($ p_i $). Each member is modeled as a Hermite cubic curve for which the end slopes ($ v^{T_1}_{m_j} $ and $ v^{T_2}_{m_j}$) and, in-plane width ($ v^w_{m_j} $) are continuous design variables that vary within the user specified limits. Each vertex $ p_i $ is permitted to translate within the specified limits by modeling its horizontal and vertical coordinates as continuous design variables. The out-of-plane thickness $ v^{th} $ of the entire continuum and magnitude of the applied input force $ v^F_I $ also form part of the continuous design variables.

A discrete topology choice variable $ v^c_{m_j} $ that can take values of either 0 or 1, decides the presence of a member in the candidate design. A candidate design shown in Fig. \ref{fig:DesignDomain}b is generated with the aforementioned design variables. It may happen that two or more members, e.g., $ m_{32} $ and $ m_{34} $ in the figure, intersect. All such intersecting members are identified and deleted by assigning 0 to the respective topology choice variables. A candidate design (Fig. \ref{fig:DesignDomain}c) after deletion of intersecting members, is meshed with quadrilateral elements. When doing so, elements overlap at the junctions. For instance, at junction $ ju_1 $, members $ m_1, m_2, m_3, \text{and } m_4 $ are not properly connected (Fig. \ref{fig:DesignDomain}d). To maintain connectivity between members at the junctions, the latter are modeled as circles of suitable radii. Further, each circular junction is meshed with quadrilateral elements as shown in  Fig. \ref{fig:DesignDomain}e. Elements at the end of members sharing the junction are connected to the circular junction. However, modeling of junctions as circles makes the junctions unnecessarily \emph{bulgy}. For better appearance, all junctions are compressed by repositioning the nodes, as detailed in \cite{Reddy-saxena-2020}. E.g., junction $ ju_1 $ after repositioning of nodes is shown in Fig. \ref{fig:DesignDomain}f. Fig.  \ref{fig:DesignDomain}g shows well connected members of the candidate design meshed with quadrilateral elements and reshaped junctions. After meshing the continuum with quadrilateral elements, we represent it as a combination of a set of inner (numbered 2 to 10) and outer (numbered 1) loops as per the procedure laid down in \cite{Reddy-saxena-2020}. Once the continuum (Fig. \ref{fig:DesignDomain}g) is ready for analysis, we specify  contact interactions as self-contact between internal and external loops individually. In addition, contact interactions are also specified between the outer loop and the workpiece to compute the output force for CoFMs and CiFMs. We use stochastic Random Mutation Hill Climber search, detailed in \cite{Reddy-saxena-2020}, for reasons mentioned in section \ref{sec:dis_cofm_cifm}.

\section{Examples}
\label{sec:to_eg_cofm}
\begin{figure*}[h!]
	\centering
	\begin{subfigure}[t]{0.32 \textwidth}
		\centering
		\includegraphics[scale=0.7]{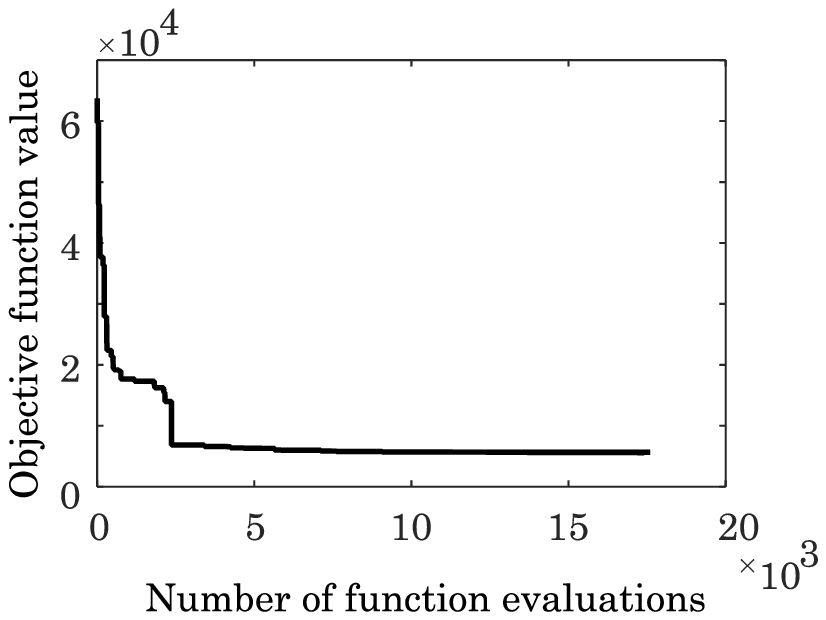}
		\caption{}
		\label{fig:Convergance_history}
	\end{subfigure} \hfill 
	\begin{subfigure}[t]{0.32 \textwidth}
		\centering
		\includegraphics[scale=.7]{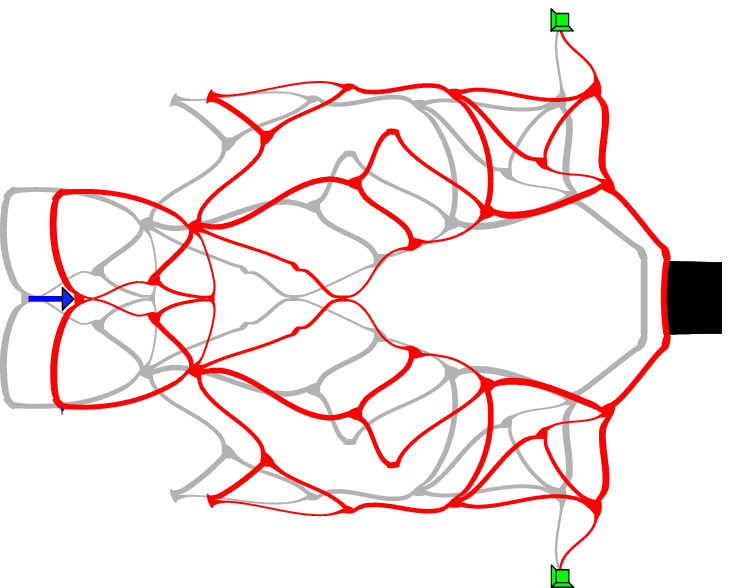}
		\caption{}
		\label{fig:CoFM_Symmetric_example1_b}
	\end{subfigure} \hfill
	\begin{subfigure}[t]{0.32 \textwidth}
		\centering
		\includegraphics[scale=.75]{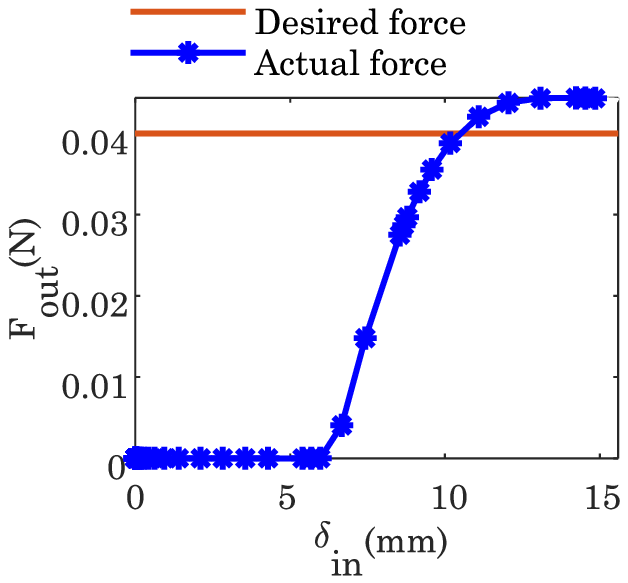}
		\caption{}
		\label{fig:CoFM_Symmetric_example1_c}
	\end{subfigure} 
	\caption{(\subref{fig:Convergance_history}) Convergence history of optimization process for CoFM-I. (\subref{fig:CoFM_Symmetric_example1_b}) Initial state (gray) of the complete design obtained after replicating and scaling the symmetric half and final (red) deformed configuration. (\subref{fig:CoFM_Symmetric_example1_c}) Force-deflection relation for CoFM-I.}
	\label{fig:Symmetric_example1}
\end{figure*}
We demonstrate proposed synthesis with two examples each, of the Constant output force (CoFM-I and CoFM-II) and Constant input force  (CiFM-I and CiFM-II) mechanisms. In all cases, we seek singlepiece continua permitting contact between only members (i.e., self contact) and that between compliant mechanisms and the respective flexible workpieces. For the four continua, additional external surfaces (to facilitate mutual contact) as in \cite{Reddy-saxena-2020} are not generated for now (see section \ref{sec:to_eg_cofm_cifm_ext_surf}). Workpieces are considered rectangular in shape and flexible with their Neo-Hookean hyper elastic material constants C10 and D1 \cite[pp. 21.5.1–5]{ABAQUS-2011} taken as 0.376 and 1.020 respectively. This corresponds to the Young's modulus of 0.02 Nmm$^{-2}$ and Poisson's ratio of 0.33. We show that the output force behavior of the synthesized continua is specific to the shape and material property of the workpieces they interact with (see section \ref{sec:f_vs_d_wp_chg}). Topology optimization is performed using the Random Mutation Hill climber (RMHC) detailed in \cite{Reddy-saxena-2020} with the optimization parameters, objective weights, member material properties and other scalars given in Table \ref{Table:Opti_parameters}. We do ensure force transfer in that we consider candidate designs where the contact between the workpiece and continuum remains active throughout the force deformation history of the continuum\footnote{The proposed method is capable of synthesizing continua with non-differentiable force deflection characteristics.}. All other designs are penalized.
{
	\begin{table}[h!]
		\centering
		\caption{Optimization parameters ---  Mutation probability: 0.08, Young's modulus (Nmm$^{-2}$): 20, 	Poisson's ratio: 0.33, Elements along  member width ($n_{ew}$): 4, 	Elements along member length ($n_{el}$): 20   }
		\label{Table:Opti_parameters}
		\begin{tabular}{c c c c c}
			\hline
			Parameter   &  CoFM-I & CoFM-II & CiFM-I & CiFM-II\\ \hline 
			Weight, slopes ($ C_1 $) & $ 10^6 $ & $10^6$ & $10^6$ & $10^8$  \\
			Weight, range ($ C_2 $) &  $10^2$ & $10^2$ & $10^3$&$10^3$  \\
			Weight, force\footnote{desired output force for CoFM, and input force for CiFM} ($ C_3 $)     & $ 10^6 $ & $10^6$ & $10^2$&$10^2$ \\	
			Weight, o/p force\footnote{only for CiFM} ($ C_4 $)   & -- & -- & -- & $10^2$\\			
			$ \lambda_{in} $     & 1.2  & 1.2 & 1 & 1 \\
			Desired o/p force ($ F^0 $)    & 0.02 N & 0.06 N & -- & -- \\
			thickness & 3mm & 3mm & 5 mm & 3 mm\\
			\hline 
		\end{tabular}
	\end{table}
}

\subsection{Constant output force mechanism CoFM-I}
\label{sec:CoFM_eg1}

We consider a domain of size 100 mm $ \times $ 100 mm to model the upper symmetric half of the mechanism along with the  workpiece, W (Fig. \ref{fig:CoFM_example1_a}). The design domain is represented with 12 blocks, 4 along the length and 3 along the width. The output port is reinforced with three additional members shown in Fig. \ref{fig:DesignDomain}a so that contact area between the workpiece and output port is increased for improved distribution of the output force. Design parameters for these additional members do not vary during optimization. Lower and upper bounds of all design variables of continuum members are given in Table \ref{Table:Range_members}. Young's Modulus and Poisson's ratio of members of the continuum are provided in Table \ref{Table:Opti_parameters}.

A candidate design is generated and mutated as per the procedure briefed in Sec. \ref{sec:to_cifm_cofm} over the optimization parameters specified in  Table \ref{Table:Opti_parameters} with an input force that varies between 0 and 5 N. The desired CoFM-I evolved in 17407 iterations is shown in Fig. \ref{fig:CoFM_example1_b}. Further, dangling members that offer no contribution to the mechanism, encircled within dashed regions, are removed to get the final design shown in  Fig. \ref{fig:CoFM_example1_c}. Final deformed configuration is shown in Fig. \ref{fig:CoFM_example1_g}. Variation of output force ($ F_{out} $) with respect to the input displacement ($ \delta_{in} $) is shown in Fig. \ref{fig:CoFM_example1_h}. One observes that the output force remains zero till the continuum comes into contact with the workpiece ($\Delta_{out} = 3.4$mm) and then starts increasing. After the continuum deforms further ($\delta_{in} \approx 7.5\text{ mm}$), the overall contact force between the vertical output member of the mechanism and the workpiece remains constant, despite continuing  actuation (increase in input force), and is very close to the desired output force of $0.02$ N. Evolution history of the optimization process is depicted in Fig. \ref{fig:Convergance_history}. Change in objective function value is significantly more in the initial iterations than later. 

The evolved design without dangling members is exported to an IGES file, which is further processed using a CAD Software to form the complete design after replicating and scaling the symmetric half as shown in Fig. \ref{fig:CoFM_Symmetric_example1_b}. A scale factor of 1.4 is used for ease in manufacturing. The entire mechanism is further analyzed using ABAQUS (Fig. \ref{fig:CoFM_Symmetric_example1_b}) and deformation history (Fig.  \ref{fig:CoFM_Symmetric_example1_c}) is extracted. One notices that the output force is more or less constant for $\delta_{in} \geq   12$ mm, and is slightly more than twice the desired value used for the continuum in Fig. \ref{fig:CoFM_example1_b}. This is expected as height of the workpiece is more than twice of that used in Fig. \ref{fig:CoFM_example1_b}, and that also, the full and scaled mechanism interacts with it.

\begin{table}[h]
	\centering
	\caption{Range of design variables for members of CoFM-I}
	\label{Table:Range_members}
	\begin{tabular}{P{0.5\textwidth} c c c}	
		\toprule 
		Design variable & LB & UB & Units\\ 
		\toprule 
		End-slopes $ (v_{m_j}^{T_1}, v_{m_j}^{T_2}) $ & $ - $0.5 & 0.5 & rad \\ 
		In-plane width $(v_{m_j}^w)$& 1.4 & 2 & mm \\ 
		Out of plane thickness $(v^{th})$& 2 & 5 & mm \\ 
		$ x $ and $ y $ coordinates of the bounding box $(v_{p_i}^x, v_{p_i}^y)$ & $ - $5 & 5 & mm \\ 
		\bottomrule 
	\end{tabular}
\end{table}

\subsection{Constant output force mechanism CoFM-II}
\label{sec:CoFM_eg2}
\begin{figure*}[t!]
	\centering
	\begin{subfigure}[t]{0.24 \textwidth}
		\centering
		\includegraphics[scale=0.6]{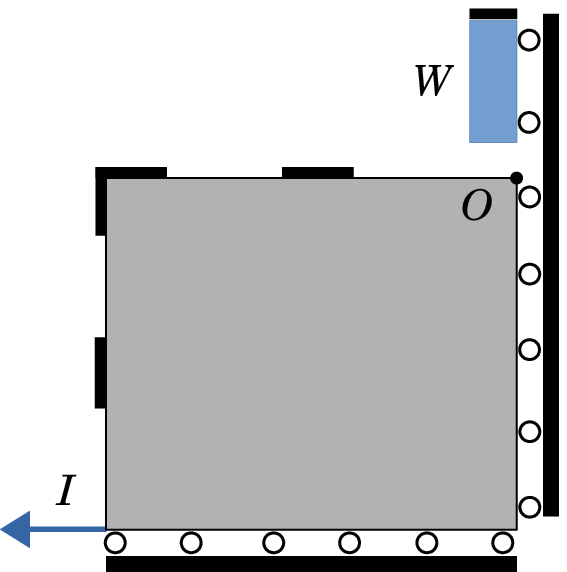}
		\caption{}
		\label{fig:CoFM_example2_a}
	\end{subfigure} 
	\begin{subfigure}[t]{0.24 \textwidth}
		\centering
		\includegraphics[scale=0.6]{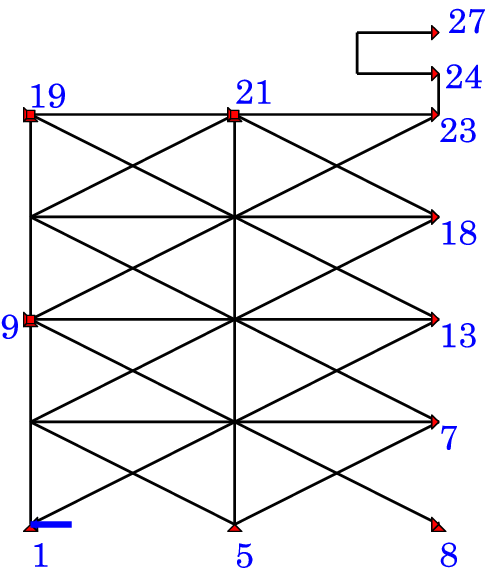}
		\caption{}
		\label{fig:CoFM_example2_b}
	\end{subfigure}
	\begin{subfigure}[t]{0.24 \textwidth}
		\centering
		\includegraphics[scale=0.6]{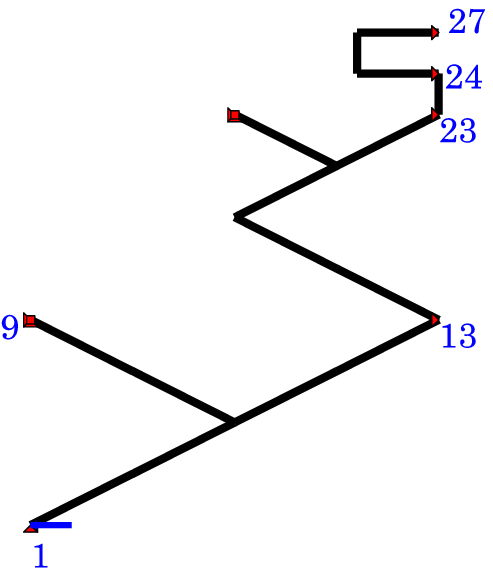}
		\caption{}
		\label{fig:CoFM_example2_c}
	\end{subfigure} 
	\begin{subfigure}[t]{0.24 \textwidth}
		\centering
		\includegraphics[scale=0.6]{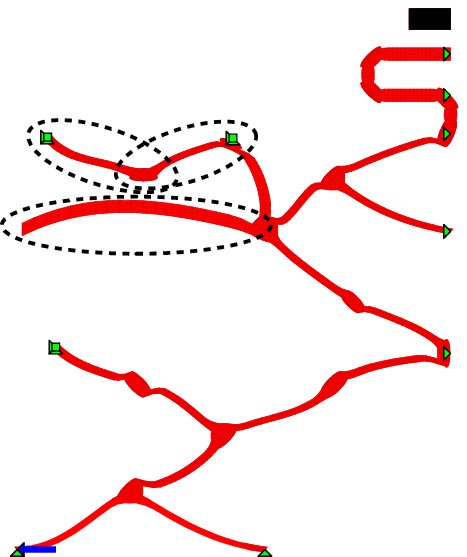}
		\caption{}
		\label{fig:CoFM_example2_d}
	\end{subfigure} \\ 
	\begin{subfigure}[t]{0.4 \textwidth}
		\centering
		\includegraphics[scale=0.75]{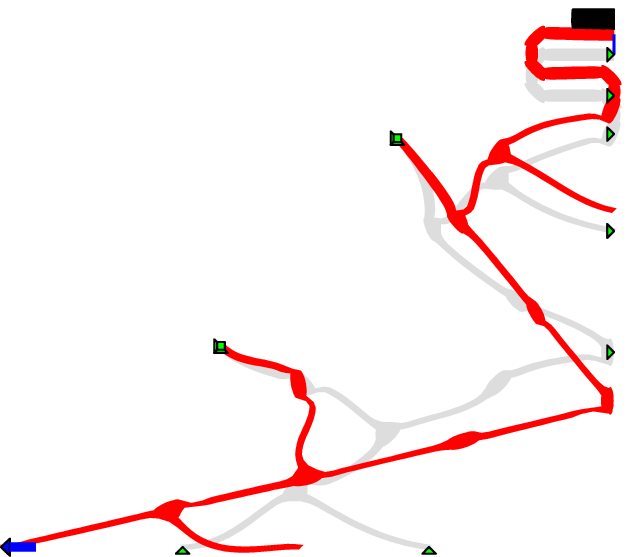}
		\caption{}
		\label{fig:CoFM_example2_h}
	\end{subfigure} 
	\begin{subfigure}[t]{0.5 \textwidth}
		\centering
		\includegraphics[scale=0.8]{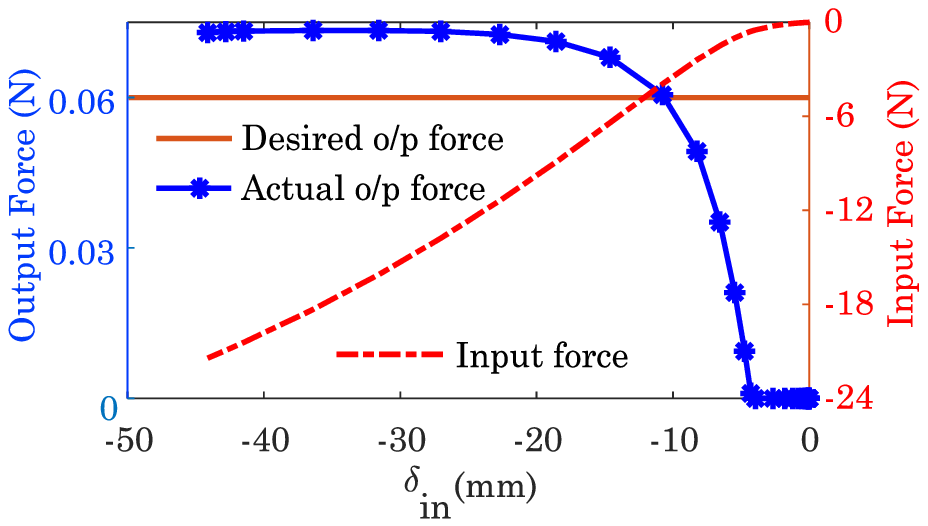}
		\caption{}
		\label{fig:CoFM_example2_i}
	\end{subfigure}
	\caption{(\subref{fig:CoFM_example2_a}) Design specifications for CoFM-II. (\subref{fig:CoFM_example2_b}) Design domain represented with frame members. (\subref{fig:CoFM_example2_c}) Initial guess for optimization as two symmetric halves of the inverter mechanism placed in series. (\subref{fig:CoFM_example2_d}) optimized design. (\subref{fig:CoFM_example2_h}) Final design after removing dangling members and final deformed configuration (red) of the symmetric half of CoFM-II. $\Delta_{\textmd{out}}:$ 4.5mm. (\subref{fig:CoFM_example2_i}) ${F_{out}} -{\delta_{in}}$ behavior of CoFM-II.}
	\label{fig:CoFM_example2}
\end{figure*}

\begin{table}[h]
	\centering
	\caption{Range of design variables for members of CoFM-2}
	\label{Table:Range_members_CoFM_eg_2}
	\begin{tabular}{P{0.5\textwidth} c c c}	
		\toprule 
		Design variable & LB & UB & Units\\ 
		\toprule 
		End-slopes $ (v_{m_j}^{T_1}, v_{m_j}^{T_2}) $ & $ - $0.5 & 0.5 & rad \\ 
		In-plane width $(v_{m_j}^w)$& 2 & 3 & mm \\ 
		Out of plane thickness $(v^{th})$& 1 & 3 & mm \\ 
		$ x $ coordinates of the bounding box $(v_{p_i}^x)$ & $ - $5 & 5 & mm \\ 
		$ y $ coordinates of the bounding box $(v_{p_i}^y)$ & $ - $2.5 & 2.5 & mm \\ 
		\bottomrule 
	\end{tabular}
\end{table}

We design CoFM-II (Fig. \ref{fig:CoFM_example2}) inspired by the inverter mechanism, wherein, the input force is applied along the negative $ x- $ direction and output is sought along the positive $ y- $ direction (Fig. \ref{fig:CoFM_example2_a}). The design being symmetric about both horizontal and vertical axes, for synthesis,  we choose one quarter of the design region of size 200 $ \times $ 200 mm, along with the workpiece ($ W $). $ I \text{ and } O $ represent the input and output ports respectively. We represent the design region with \textit{eight} blocks, \textit{two} along the length and \textit{four} along the width (Fig. \ref{fig:CoFM_example2_b}). The mesh comprises of 48 candidate frame members and 23 vertices. Additionally, four members are added to the output port as reinforcement for improved force transfer, similar to the CoFM-I example in Sec. \ref{sec:CoFM_eg1}. Material properties of the continuum and the workpiece are respectively identical to those in the previous example.

We commence optimization with the initial guess shown in \ref{fig:CoFM_example2_c} which is constituted of the symmetric halves of two inverters arranged in series. Intermediate candidate designs are mutated over the parameters given in Table \ref{Table:Opti_parameters} and allowable limits of all continuum design variables are mentioned in Table  \ref{Table:Range_members_CoFM_eg_2}. With the same input force range as that for CoFM-I, an acceptable design  shown in Fig. \ref{fig:CoFM_example2_d} is achieved. The final design (undeformed configuration in grey) obtained by removing the dangling members that do not store any energy at any stage in the deformation history is shown in Fig. \ref{fig:CoFM_example2_h}. Comparing the initial guess and this solution, there are few additional members in the latter. Final deformed configuration of the symmetric half of CoFM-II is shown in red. ${F_{out}} -{\delta_{in}}$ behavior is depicted in Fig. \ref{fig:CoFM_example2_i}. One observes that constant output force is exerted by the continuum over a range starting from a stage with $\delta_{in} \approx 20\text{ mm}$. One also observes that this is despite the input force/displacement increasing consistently. 

\begin{figure*}[h!]
	\centering
	\begin{subfigure}[t]{0.5 \textwidth}
		\centering
		\includegraphics[scale=1]{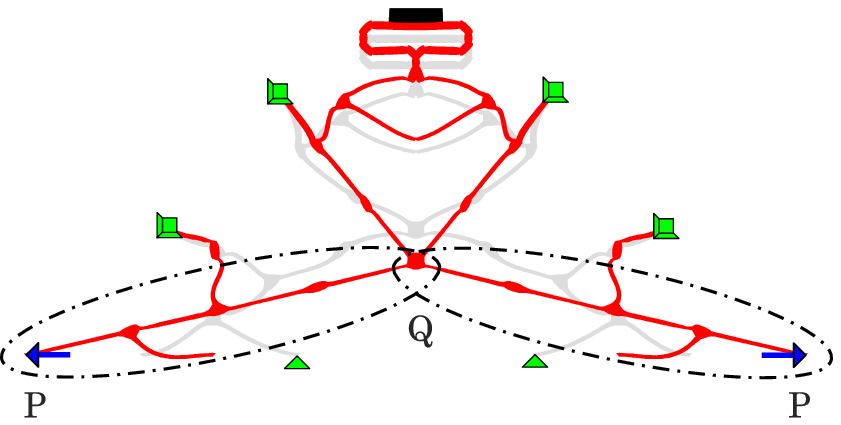}
		\caption{}
		\label{fig:CoFM_Symmetric_example2_b}
	\end{subfigure}
	\begin{subfigure}[t]{0.39 \textwidth}
		\centering
		\includegraphics[scale=0.9]{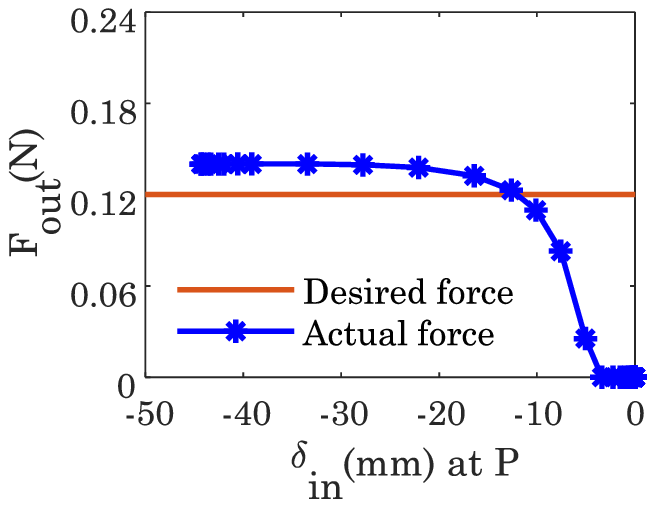}
		\caption{}
		\label{fig:CoFM_Symmetric_example2_c}
	\end{subfigure}
	\begin{subfigure}[t]{0.5 \textwidth}
		\centering
		\includegraphics[scale=0.85]{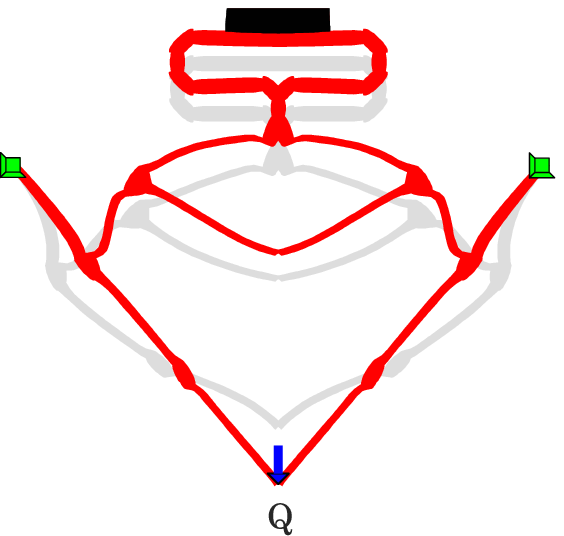}
		\caption{}
		\label{fig:CoFM_Symmetric_example2_e}
	\end{subfigure}
	\begin{subfigure}[t]{0.39 \textwidth}
		\centering
		\includegraphics[scale=1]{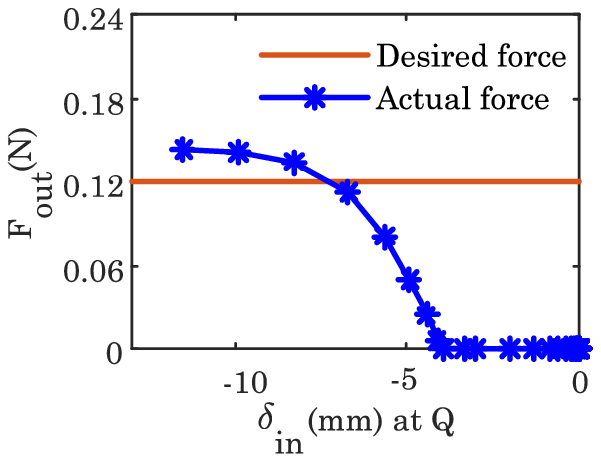}
		\caption{}
		\label{fig:CoFM_Symmetric_example2_f}
	\end{subfigure}
	\caption{
		(\subref{fig:CoFM_Symmetric_example2_b}) Initial (gray) and final (red) deformed configuration of CoFM-II. (\subref{fig:CoFM_Symmetric_example2_c}) Force-deflection behavior. $\delta_{in}$ corresponds to the horizontal displacement of point P. (\subref{fig:CoFM_Symmetric_example2_e}) Compact CoFM-II (gray), after removing members within the dashed regions in Fig. (\ref{fig:CoFM_Symmetric_example2_b}) and its deformed (red) configuration.  (\subref{fig:CoFM_Symmetric_example2_f}) ${F_{out}} -{\delta_{in}}$ behavior of the modified CoFM-II. $\delta_{in}$ corresponds to the downward vertical deflection of the bottommost (new input) node in (\subref{fig:CoFM_Symmetric_example2_e}).}
	\label{fig:Symmetric_example2_1}
\end{figure*}

To simulate the behavior of upper symmetric half of CoFM-II, we consider  replica of the evolved quarter design about the vertical axis and generate the CAD drawing, which is further analyzed in ABAQUS and the nodal displacements are extracted. Fig. \ref{fig:CoFM_Symmetric_example2_b} depict the undeformed and deformed configurations of CoFM-II respectively. The overall deflection characteristics are shown in Fig. \ref{fig:CoFM_Symmetric_example2_c}. Members within the dashed region in Fig. \ref{fig:CoFM_Symmetric_example2_b} deform significantly and stress values are as high as 22.29 Nmm$^{-2}$. We note further that a similar force behavior at the output is possible if dashed members are removed and the remnant continuum is actuated vertically downward at point Q as shown. A more concised CoFM-II and its deformed profile is shown in Fig. \ref{fig:CoFM_Symmetric_example2_e}, and the corresponding force-deflection behavior in Fig. \ref{fig:CoFM_Symmetric_example2_f}. 


\section{Constant input force mechanisms}
\label{sec:to_eg_cifm}

\begin{figure*}[h!]
	\centering
	\begin{subfigure}[t]{0.32 \textwidth}
		\centering
		\includegraphics[scale=0.75]{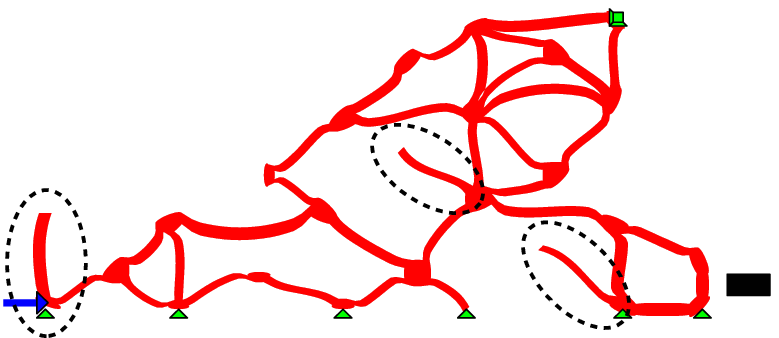}
		\caption{}
		\label{fig:CiFM_example1_a}
	\end{subfigure}\hfill
	\begin{subfigure}[t]{0.32 \textwidth}
		\centering
		\includegraphics[scale=0.7]{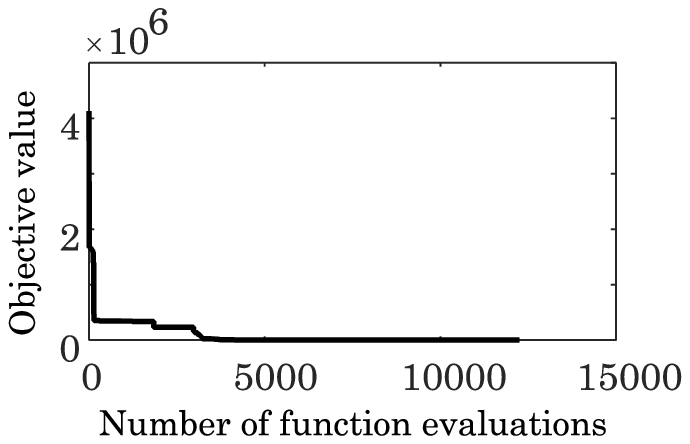}
		\caption{}
		\label{fig:CiFM_example1_b}
	\end{subfigure}
	\begin{subfigure}[t]{0.32 \textwidth}
		\centering
		\includegraphics[scale=0.75]{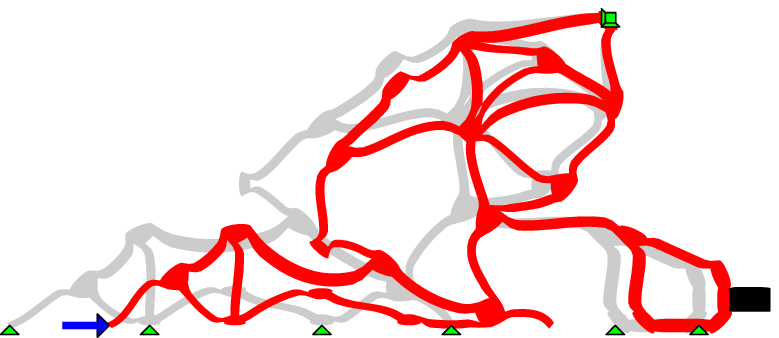}
		\caption{}
		\label{fig:CiFM_example1_c}
	\end{subfigure}\\
	\begin{subfigure}[t]{0.32 \textwidth}
		\centering
		\includegraphics[scale=0.7]{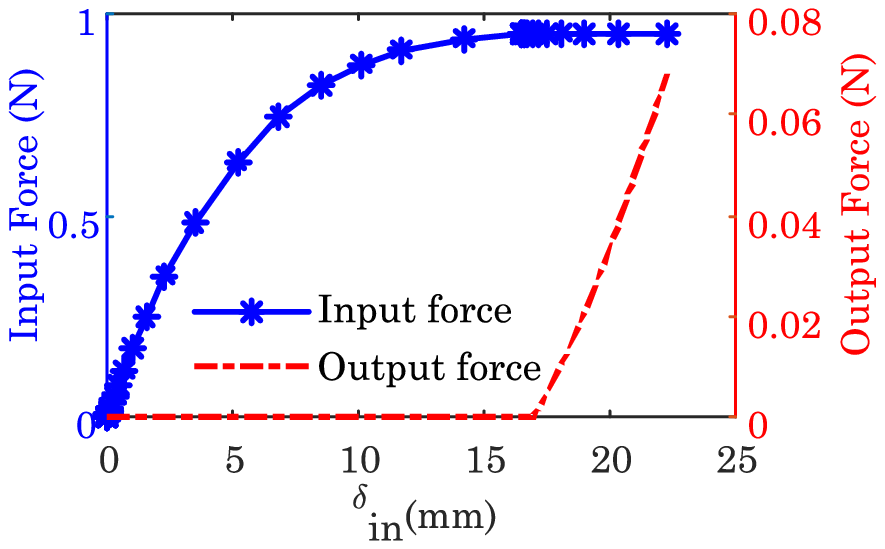}
		\caption{}
		\label{fig:CiFM_example1_d}
	\end{subfigure}
	\begin{subfigure}[t]{0.32 \textwidth}
		\centering
		\includegraphics[scale=0.65]{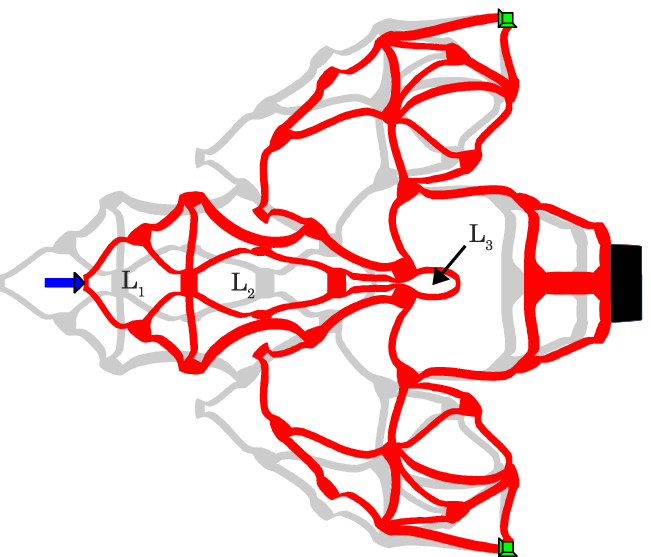}
		\caption{}
		\label{fig:CiFM_example1_f}
	\end{subfigure}
	\begin{subfigure}[t]{0.32 \textwidth}
		\centering
		\includegraphics[scale=0.7]{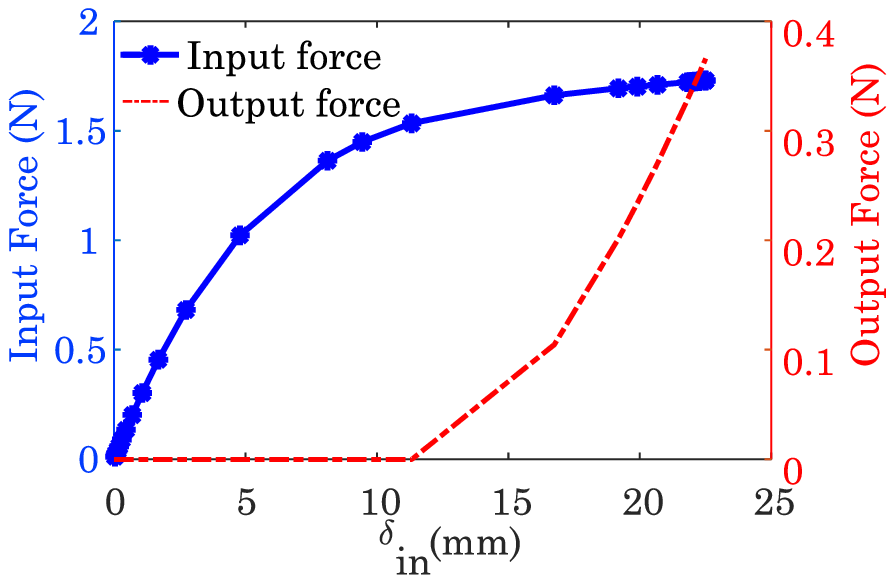}
		\caption{}
		\label{fig:CiFM_example1_g}
	\end{subfigure}\\
	\caption{(\subref{fig:CiFM_example1_a}) Constant input Force Mechanism (CiFM-I) synthesized using the design specifications in Fig. \ref{fig:CoFM_example1_a}. (\subref{fig:CiFM_example1_b}) Synthesis takes about 12000 evaluations. (\subref{fig:CiFM_example1_c}) Upper symmetric half of CiFM-I with dangling elements (e.g., encircled portions in (\subref{fig:CiFM_example1_a}) removed.  $\Delta_{\textmd{out}}:$ 4.6mm. (\subref{fig:CiFM_example1_d}) Force-deflection history-- input force is constant as expected. (\subref{fig:CiFM_example1_f}) Extensions of 5 mm used between the symmetric halves of the full continuum. (\subref{fig:CiFM_example1_g}) Corresponding force-deflection history. The input force is almost doubled, and the output force is increased significantly due to a larger workpiece used. }
	\label{fig:CiFM_example1}
\end{figure*}

We perform continuum optimization for Constant input Force Mechanisms (CiFMs- I and II) with identical design specifications as for CoFM-I (Fig. \ref{fig:CoFM_example1_a}) and slightly revised specifications as for CoFM-II (Fig. \ref{fig:CoFM_example2_a}) respectively. The focus in these examples is to have the input force, as opposed to the output force,  unvarying over a range of input displacements. Objective in Eq. \ref{lc_CiFM_obj} is minimized with the weights given in Table. \ref{Table:Opti_parameters}. $C_4$ is set to zero in that the output force (Eq. \ref{lc_CiFM_obj}) is not maximized.

Fig. \ref{fig:CiFM_example1_a} shows the solution for the upper symmetric half of CiFM-I obtained in about 12500 function evaluations (Fig. \ref{fig:CiFM_example1_b}). The cleaned-up solution, after removal of dangling sub-regions, and its deformed profile are shown in Fig. \ref{fig:CiFM_example1_c}. The corresponding forces vs. input displacement curves are depicted in Fig. \ref{fig:CiFM_example1_d}. As intended, the input force is almost constant, and close to $1$ N, over the input displacement range between $15-22$ mm. Within the same range, the output force increases till $0.07$ N ensuring force transfer throughout the mechanism actuation. The solutions in Figs. \ref{fig:CoFM_example1_c} and \ref{fig:CiFM_example1_c}, although synthesized using the same specifications but with different design intent, are different, and so are their respective functional responses in Figs. \ref{fig:CoFM_example1_h} and \ref{fig:CiFM_example1_d}. This reinforces that constant input force mechanisms need not be constant output force mechanisms, and vice versa. Evidently, they are two different categories of constant force mechanisms.

Fig. \ref{fig:CiFM_example1_f} shows full mechanism wherein the upper symmetric half is replicated about the horizontal axis. The gap between the two halves is set to $5$ mm. Self contact is modeled for the three central loops $L_1, L_2$ and $L_3$ shown in the Fig. \ref{fig:CiFM_example1_f}. During synthesis, such contact was not modeled between involved members and the horizontal axis. Extensions\footnote{These additions are expected to influence the overall stiffness and hence the deformation characteristics} are introduced via thick regions at locations where the nodes (green triangles in Fig. \ref{fig:CiFM_example1_c}) are set on rollar support. Force deflection relation is shown in Fig. \ref{fig:CiFM_example1_g}. Slope of the input force vs. the input displacement curve is almost flat especially for $\delta_{in} >17$ mm and not quite close to zero.

\begin{figure*}[h!]
	\centering
	\begin{subfigure}[t]{0.32 \textwidth}
		\centering
		\includegraphics[scale=0.75]{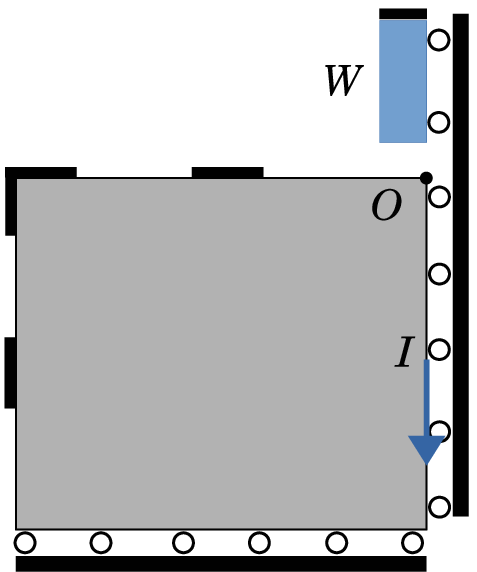}
		\caption{}
		\label{fig:CiFM_example2_a}
	\end{subfigure}\hfill
	\begin{subfigure}[t]{0.32 \textwidth}
		\centering
		\includegraphics[scale=0.6]{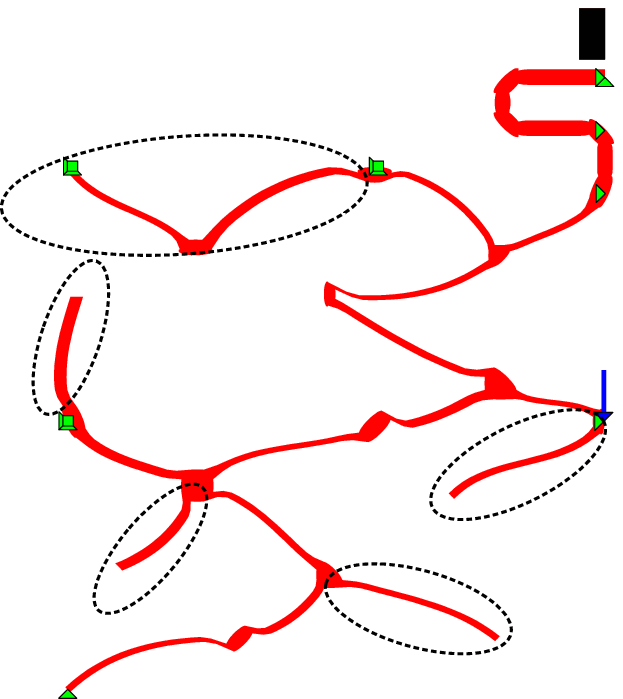}
		\caption{}
		\label{fig:CiFM_example2_d}
	\end{subfigure}
	\begin{subfigure}[t]{0.32 \textwidth}
		\centering
		\includegraphics[scale=0.6]{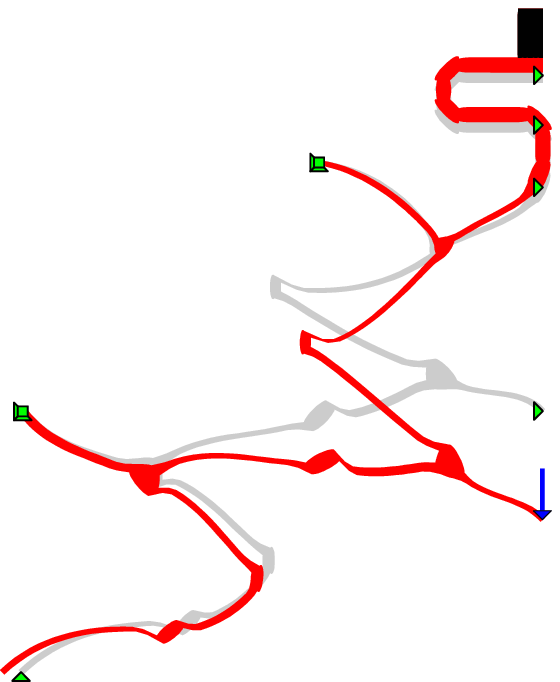}
		\caption{}
		\label{fig:CiFM_example2_e}
	\end{subfigure}
	\begin{subfigure}[t]{0.32 \textwidth}
		\centering
		\includegraphics[scale=0.6]{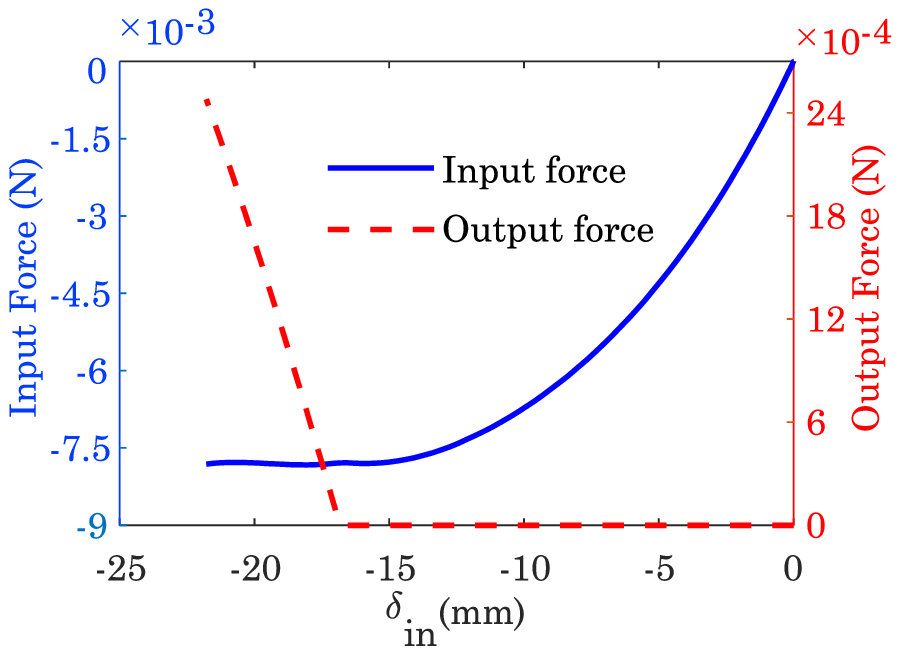}
		\caption{}
		\label{fig:CiFM_example2_f}
	\end{subfigure}
	\begin{subfigure}[t]{0.32 \textwidth}
		\centering
		\includegraphics[scale=0.6]{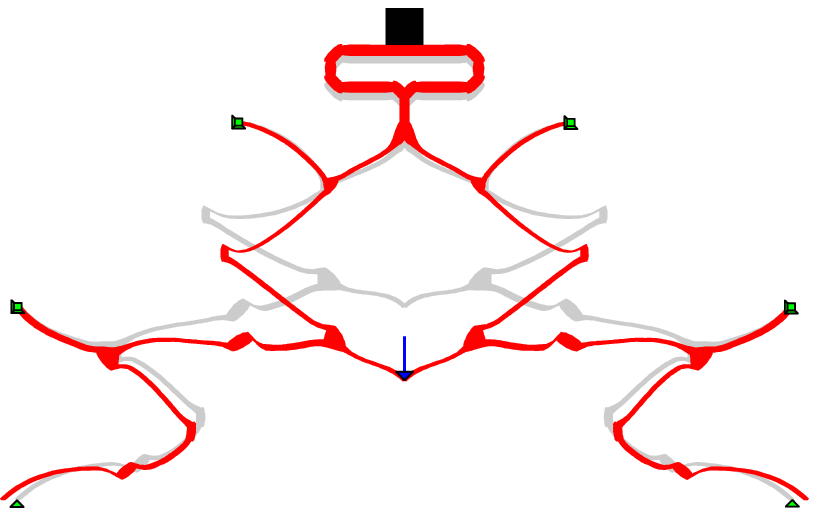}
		\caption{}
		\label{fig:CiFM_example2_symmetric_a}
	\end{subfigure}
	\begin{subfigure}[t]{0.32 \textwidth}
		\centering
		\includegraphics[scale=0.65]{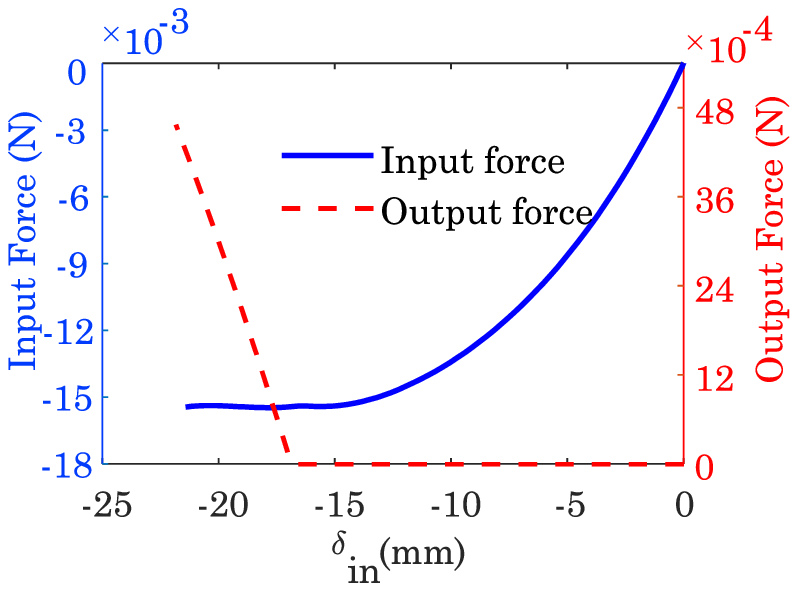}
		\caption{}
		\label{fig:CiFM_example2_symmetric_b}
	\end{subfigure}
	\label{fig:CiFM_example2}
	\caption{Constant input force mechanism, CiFM-II.  (\subref{fig:CiFM_example2_a}) Design specifications altered slightly compared to those in Fig. \ref{fig:CoFM_example2_a}. (\subref{fig:CiFM_example2_d}) Synthesized left symmetric half and (\subref{fig:CiFM_example2_e}) the same without members storing no internal energy. $\Delta_{\textmd{out}}:$ 1.9mm. (\subref{fig:CiFM_example2_f}) Force deflection history. (\subref{fig:CiFM_example2_symmetric_a}) full upper half of CiFM-II, which by itself is fine, if the bottom two nodes are on rollar support (\subref{fig:CiFM_example2_symmetric_b}) force deflection characteristics for (\subref{fig:CiFM_example2_symmetric_a}). }
\end{figure*}


Refering to the design specifications for CoFM-II (Fig. \ref{fig:CoFM_example2_a}), the obtained processed design and its final deformed configuration (Fig. \ref{fig:CoFM_example2_h}), we note that portion of the continuum undergoes  significant deformation (Fig. \ref{fig:CoFM_Symmetric_example2_b}). Alternatively, certain revisions can be made to the location and direction of the input force (Fig. \ref{fig:CoFM_Symmetric_example2_e}) to get a more compact design without loss in functionality. The corresponding changes in the location and direction of the input force are made (compare Figs. \ref{fig:CoFM_example2_a} and \ref{fig:CiFM_example2_a}), and CiFM-II is synthesized with the intent of keeping the input force constant over a range of input displacement while maximizing force transfer at the output. The domain and initial guess are taken the same as in Figs. \ref{fig:CoFM_example2_b} and  \ref{fig:CoFM_example2_c} respectively. Design obtained after around 25000 evaluations is depicted in Fig. \ref{fig:CiFM_example2_d} and the processed continuum, one after removing all dangling elements is shown in Fig. \ref{fig:CiFM_example2_e} in its final configuration. The input force (Fig. \ref{fig:CiFM_example2_f}) is observed to be constant and close to $7.5 \times 10^{-3}$ N is magnitude. Fig. \ref{fig:CiFM_example2_symmetric_a} shows the full mechanism with its force deflection characteristic in Fig. \ref{fig:CiFM_example2_symmetric_b}. Magnitude of the constant input force is almost doubled in comparison to the symmetric counterpart, as expected. Further, this magnitude is quite less suggesting the continuum could potentially be employed as (the building block of) a statically balanced mechanism (see Ref. \cite{Tolman-et-al-2016}).

\section{Constant output and Constant input force mechanisms with external surfaces}
\label{sec:to_eg_cofm_cifm_ext_surf}

\begin{figure*}[h!]
	\centering
	\begin{subfigure}[t]{0.3\textwidth}
		\centering
		\includegraphics[scale=0.75]{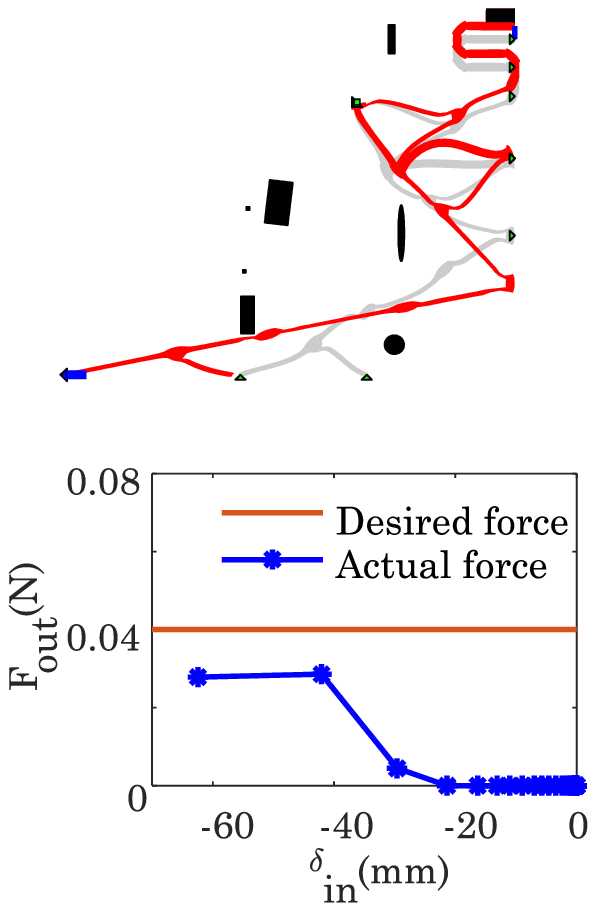}
		\caption{CoFM-II-E}
		\label{fig:CoFM-IIe}
	\end{subfigure}
	\begin{subfigure}[t]{0.36\textwidth}
		\centering
		\includegraphics[scale=0.75]{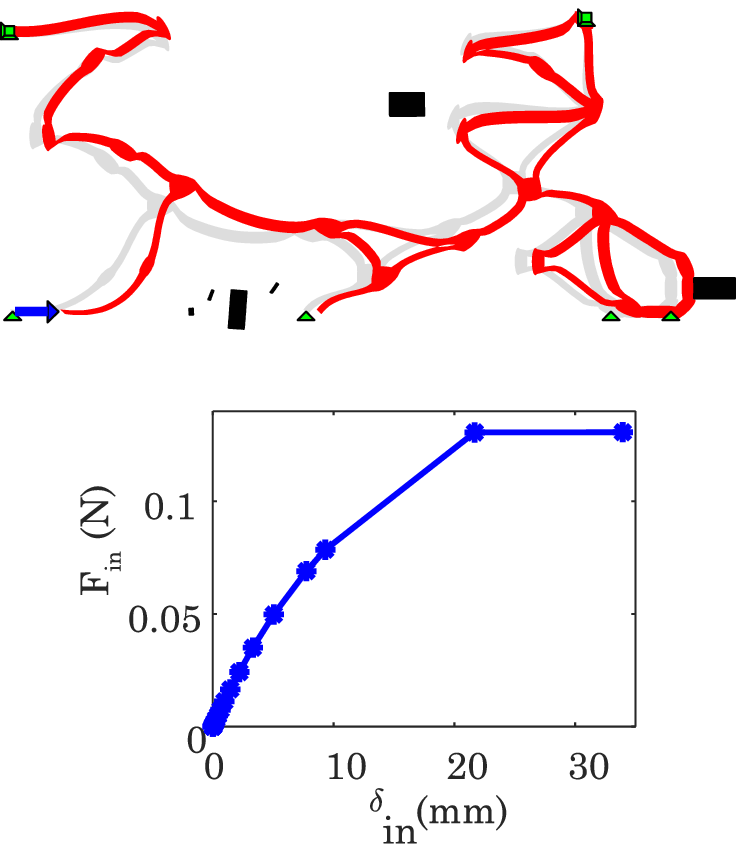}
		\caption{CiFM-I-E}
		\label{fig:CiFM-Ie}
	\end{subfigure}
	\begin{subfigure}[t]{0.3\textwidth}
		\centering
		\includegraphics[scale=0.77]{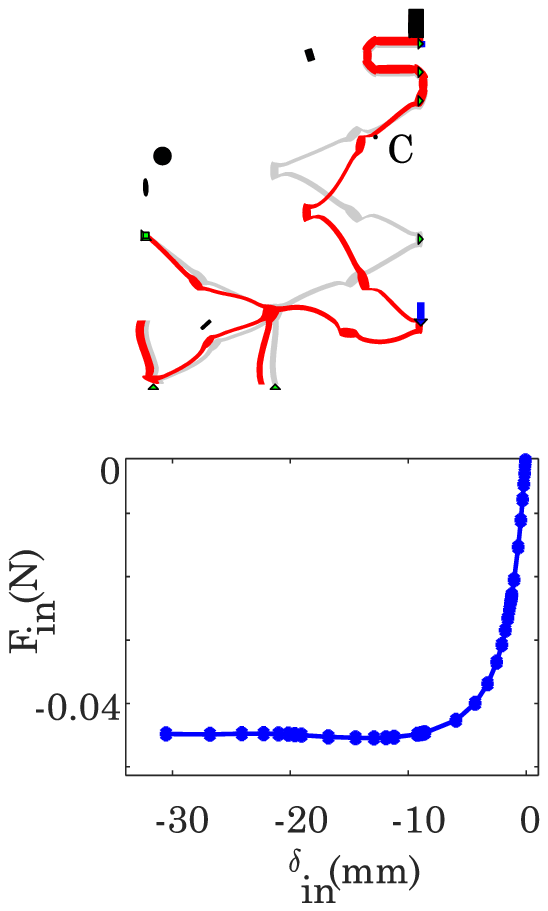}
		\caption{CiFM-II-E}
		\label{fig:CiFM-IIe}
	\end{subfigure} 
	\caption{(top) Symmetric halves of the CoFMs and CiFMs designed with self (continuum members interacting with themselves) and mutual (continuum members interacting with external, black regions) contact. 'E' represents external surfaces. (bottom)  respective force deflection curves. Self contact occurs in CiFM-II-E, and mutual contact in CiFM-II-E.}		
	\label{fig:CFMs_External_surfaces}
\end{figure*}

To determine whether permitting presence of external surfaces has any impact on the design process, we synthesize the Constant input and Constant output force mechanisms again, this time, permitting mutual contact as well. A maximum of 11 external contact surfaces of various shapes, e.g., rectangular, circular and elliptical are permitted. Symmetric halves of CoFM-II-E (i.e. CoFM-II with external surfaces) synthesized using the specifications in Fig. \ref{fig:CoFM_example2_a}, CiFM-I-E, and CiFM-II-E, designed respectively using specifications in Figs. \ref{fig:CoFM_example1_a} and \ref{fig:CiFM_example2_a}, are shown in Fig. \ref{fig:CFMs_External_surfaces}. In case of CoFM-II-E (Fig. \ref{fig:CoFM-IIe}) and CiFM-I-E (Fig. \ref{fig:CiFM-Ie}), the respective continua throughout their deformation history do not come into contact with any external surface present in their vicinity. An instance of self contact is observed in CoFM-II-E. In case of CiFM-II-E (Fig. \ref{fig:CiFM-IIe}), the continuum interacts with a small circular surface at location $C$ which is when the direction of motion of the output port changes from downward to upward. Permitting mutual contact, may not necessarily yield better, or faster solutions. In our experience, these solutions were obtained after many design evaluations. Many intermediate solutions may have been penalized due to external surfaces overlapping with the parent continuum. However, interesting alternate solutions, e.g., CiFM-II-E, may be possible if contact with external surfaces is permitted in topology synthesis.

\section{Discussion}
\label{sec:dis_cofm_cifm}
Modeling contact in synthesis of compliant mechanisms renders the latter special deformation characteristics, e.g., traversing of non-smooth paths (\cite{Nilesh_Suresh_2007, kumar2016synthesis, kumar2019computational, Reddy-saxena-2020}) and stress relief (\cite{mehta2009stress}). In this paper, we incorporate contact in synthesis to achieve constant input and output force characteristics. Besides exploring interesting design possibilities, contact analysis is specifically necessitated to truly capture interactive forces between the continua and the respective hyper flexible workpieces. With the two  synthesized CoFMs (Figs. \ref{fig:CoFM_Symmetric_example1_b} and \ref{fig:CoFM_Symmetric_example2_e}) and CiFMs (Figs. \ref{fig:CiFM_example1_f} and \ref{fig:CiFM_example2_symmetric_a}), herein, we determine on how they perform when the properties and shape of the workpiece are altered. Thereafter, we comment on the nature of search, time taken and the need for why such a search algorithm is chosen. We finally fabricate our designs using a material different from what was used to synthesize them, and test whether constant force characteristics can still be retained.

\subsection{Effect on the force-deflection curves with change in  shape/properties of the workpiece}
\label{sec:f_vs_d_wp_chg}

\begin{figure*}[h!]
	\centering
	\begin{subfigure}[t]{0.23 \textwidth}
		\centering
		\includegraphics[scale=0.3]{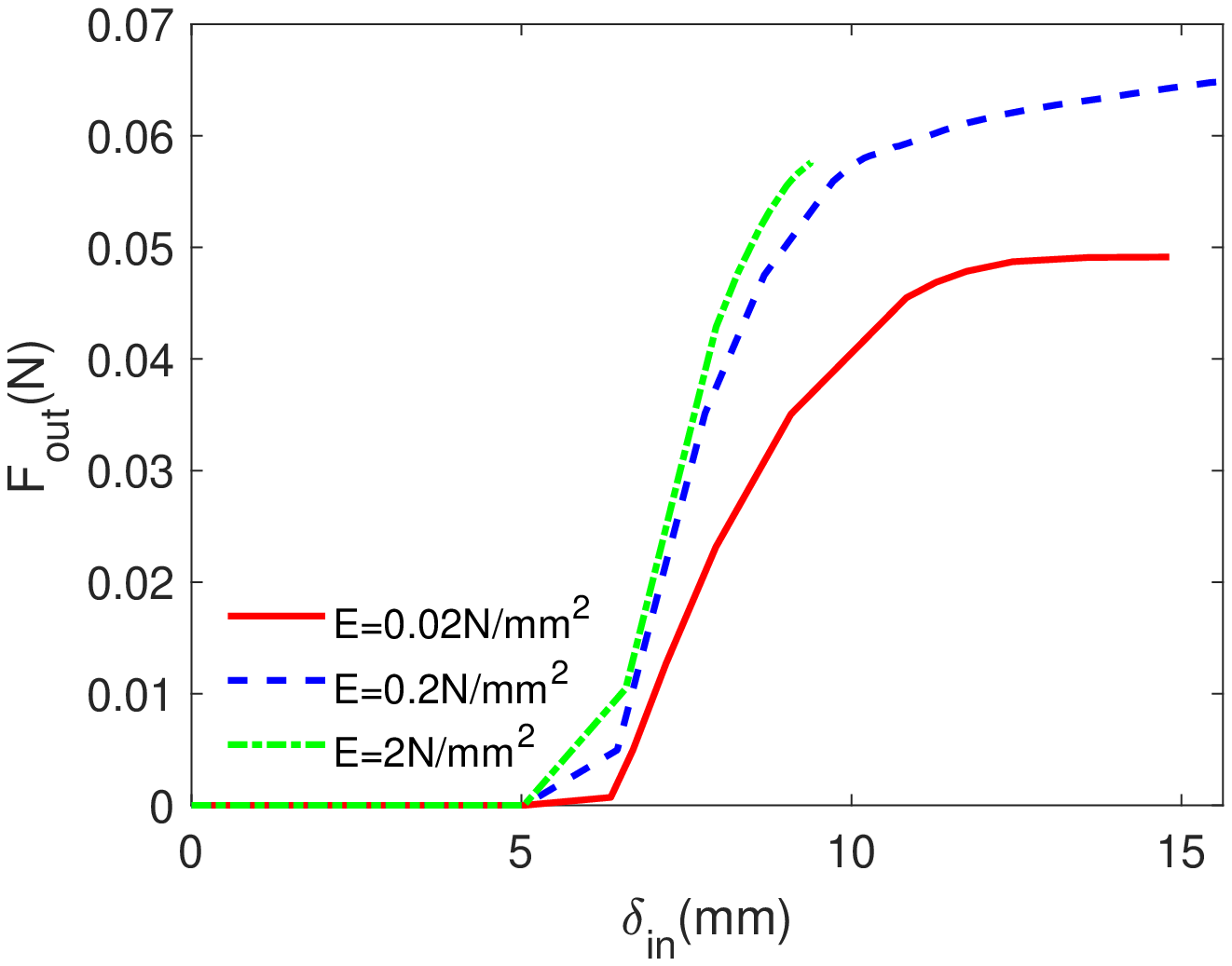}
		\caption{CoFM-I}
		\label{fig:CoFM1_wps_rectangular_various_Es}
	\end{subfigure} 
	\begin{subfigure}[t]{0.23 \textwidth}
		\centering
		\includegraphics[scale=0.3]{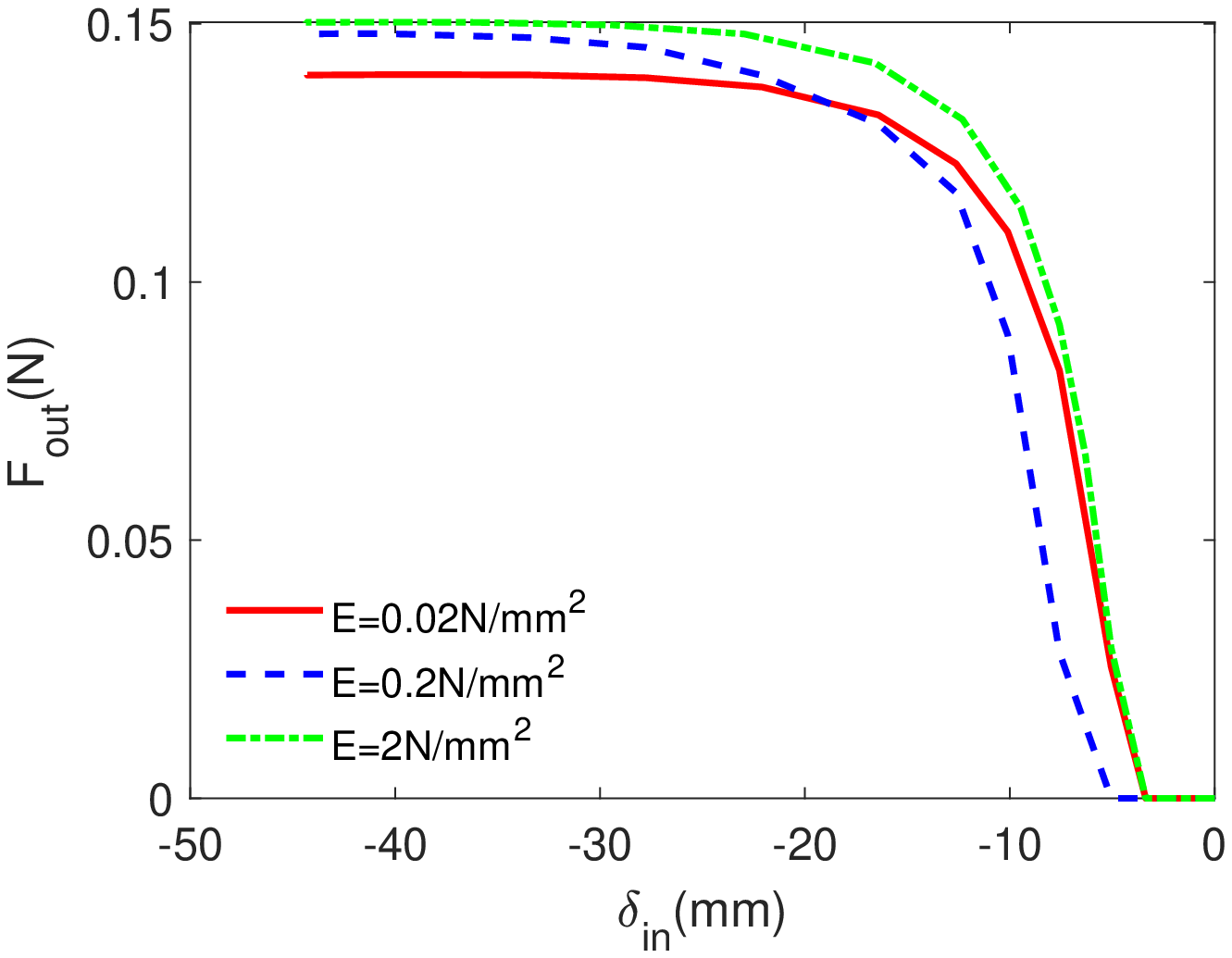}
		\caption{CoFM-II}
		\label{fig:CoFM2_wps_rectangular_various_Es}
	\end{subfigure}	
	\begin{subfigure}[t]{0.23 \textwidth}
		\centering
		\includegraphics[scale=0.3]{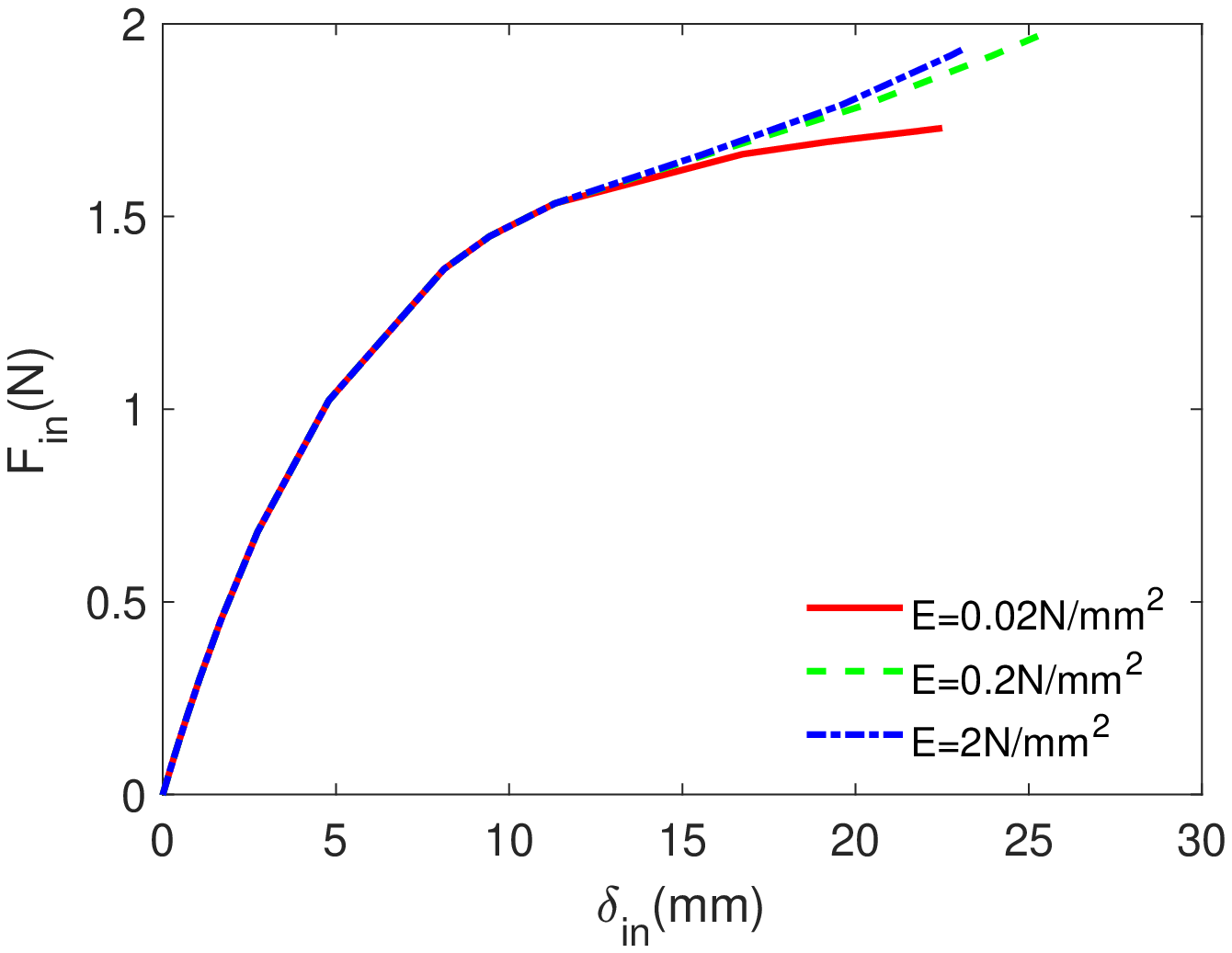}
		\caption{CiFM-I}
		\label{fig:CiFM1_wps_rectangular_various_Es}
	\end{subfigure}  
	\begin{subfigure}[t]{0.23 \textwidth}
		\centering
		\includegraphics[scale=0.3]{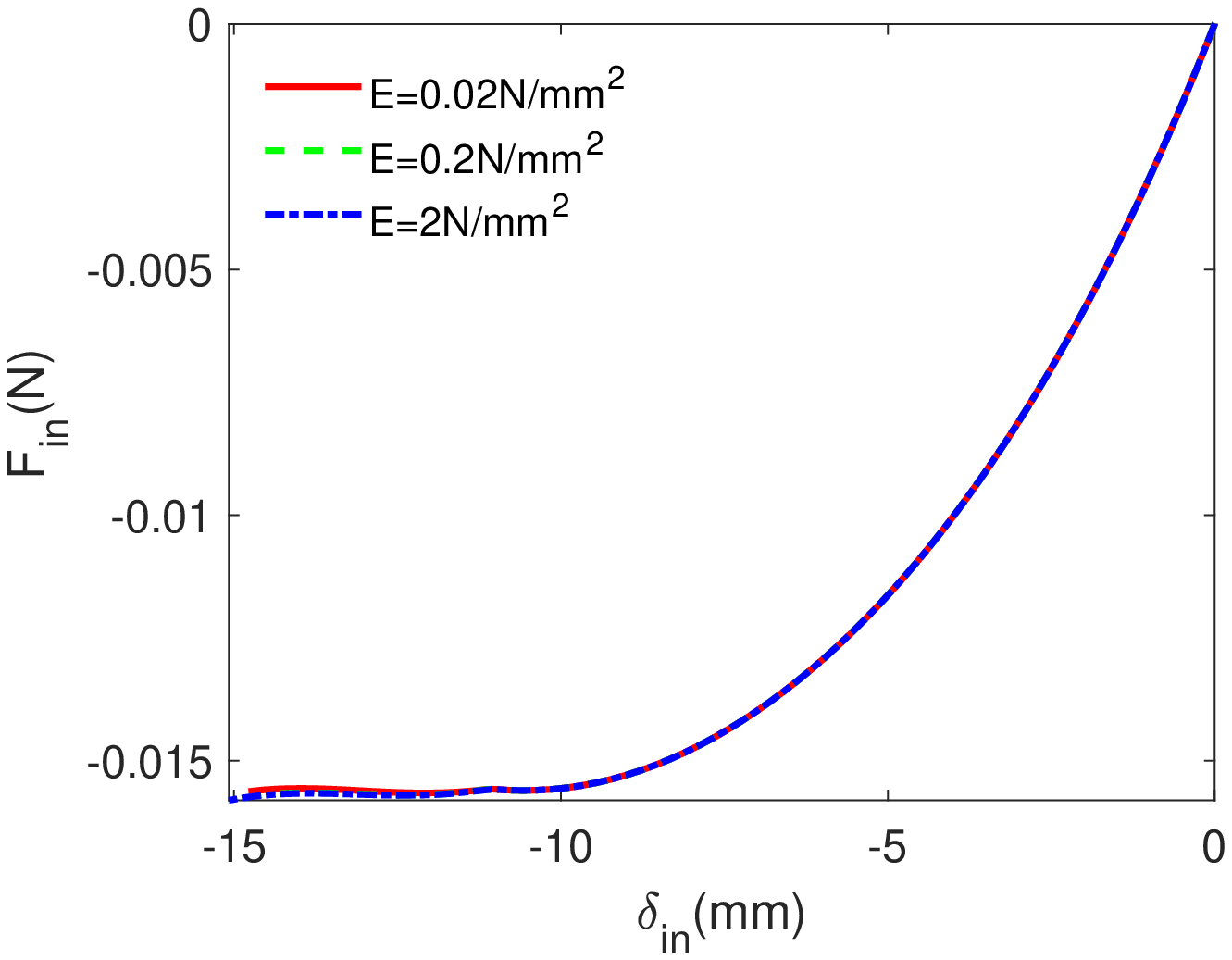}
		\caption{CiFM-II}
		\label{fig:CiFM2_wps_rectangular_various_Es}
	\end{subfigure} 
	\caption{Force-deflection characteristics for (\subref{fig:CoFM1_wps_rectangular_various_Es}) CoFM-I, (\subref{fig:CoFM2_wps_rectangular_various_Es}) CoFM-II, (\subref{fig:CiFM1_wps_rectangular_various_Es}) CiFM-I and   
		(\subref{fig:CiFM2_wps_rectangular_various_Es}) CiFM-II when the elastic modulus of the workpiece is changed. Shape of the workpiece remains rectangular, and size remains the same. }
	\label{fig:CFMs_f_d_curves_rectangular_wp_props}
\end{figure*}

\begin{figure*}[h!]
	\centering
	\begin{subfigure}[t]{0.23 \textwidth}
		\centering
		\includegraphics[scale=0.3]{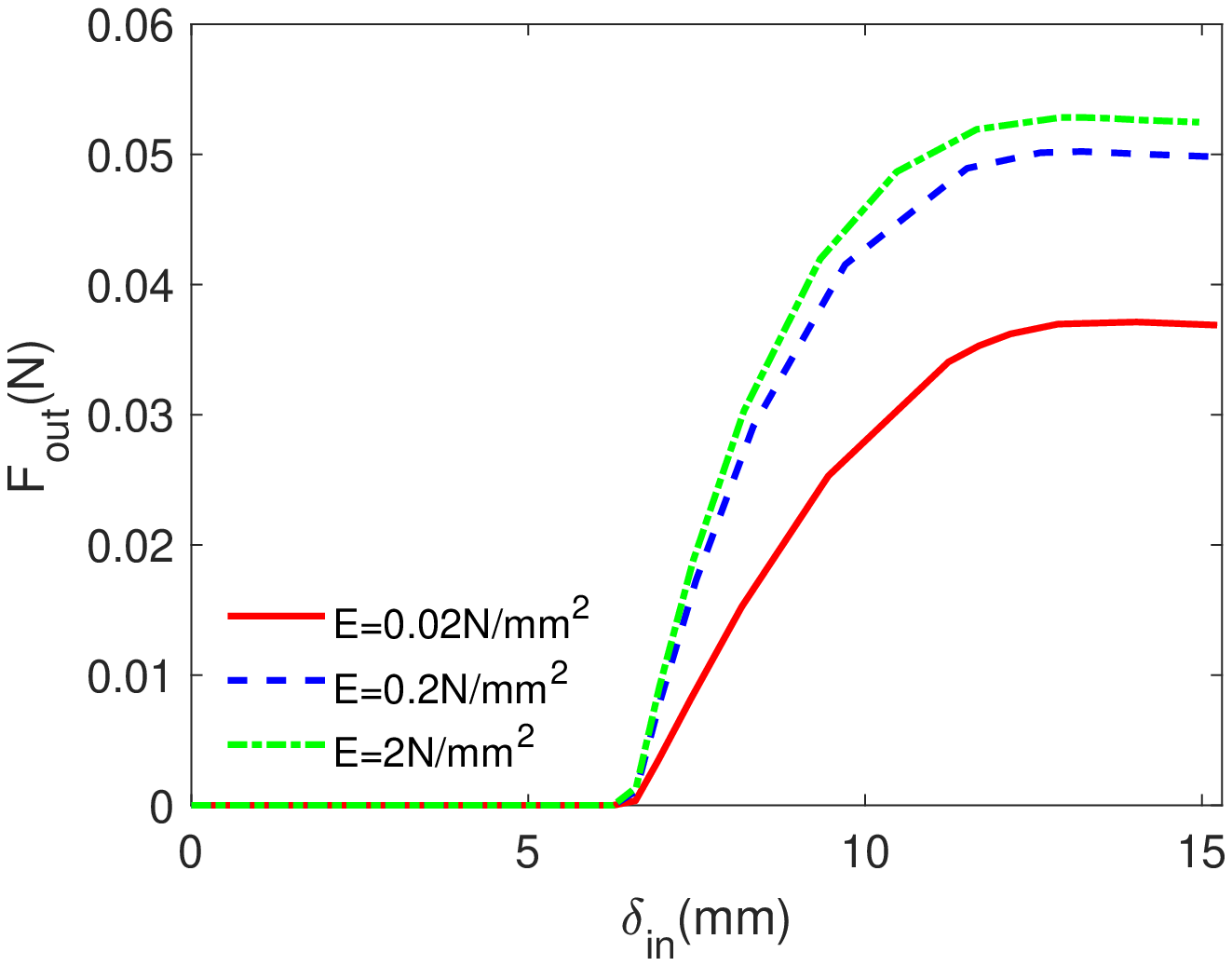}
		\caption{CoFM-I}
		\label{fig:CoFM1_wps_elliptical_various_Es}
	\end{subfigure}  
	\begin{subfigure}[t]{0.23 \textwidth}
		\centering
		\includegraphics[scale=0.3]{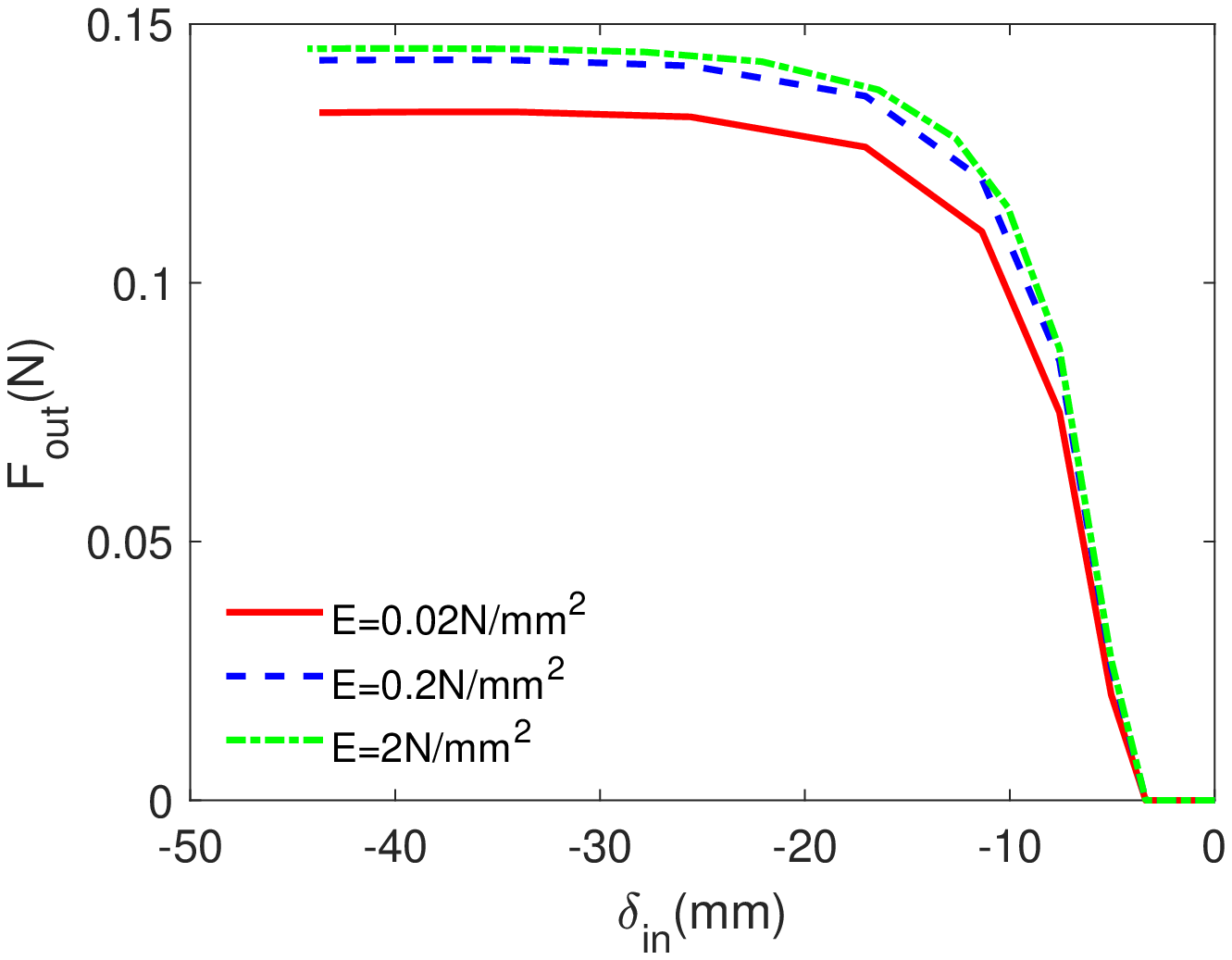}
		\caption{CoFM-II}
		\label{fig:CoFM2_wps_elliptical_various_Es}
	\end{subfigure} 
	\begin{subfigure}[t]{0.23 \textwidth}
		\centering
		\includegraphics[scale=0.3]{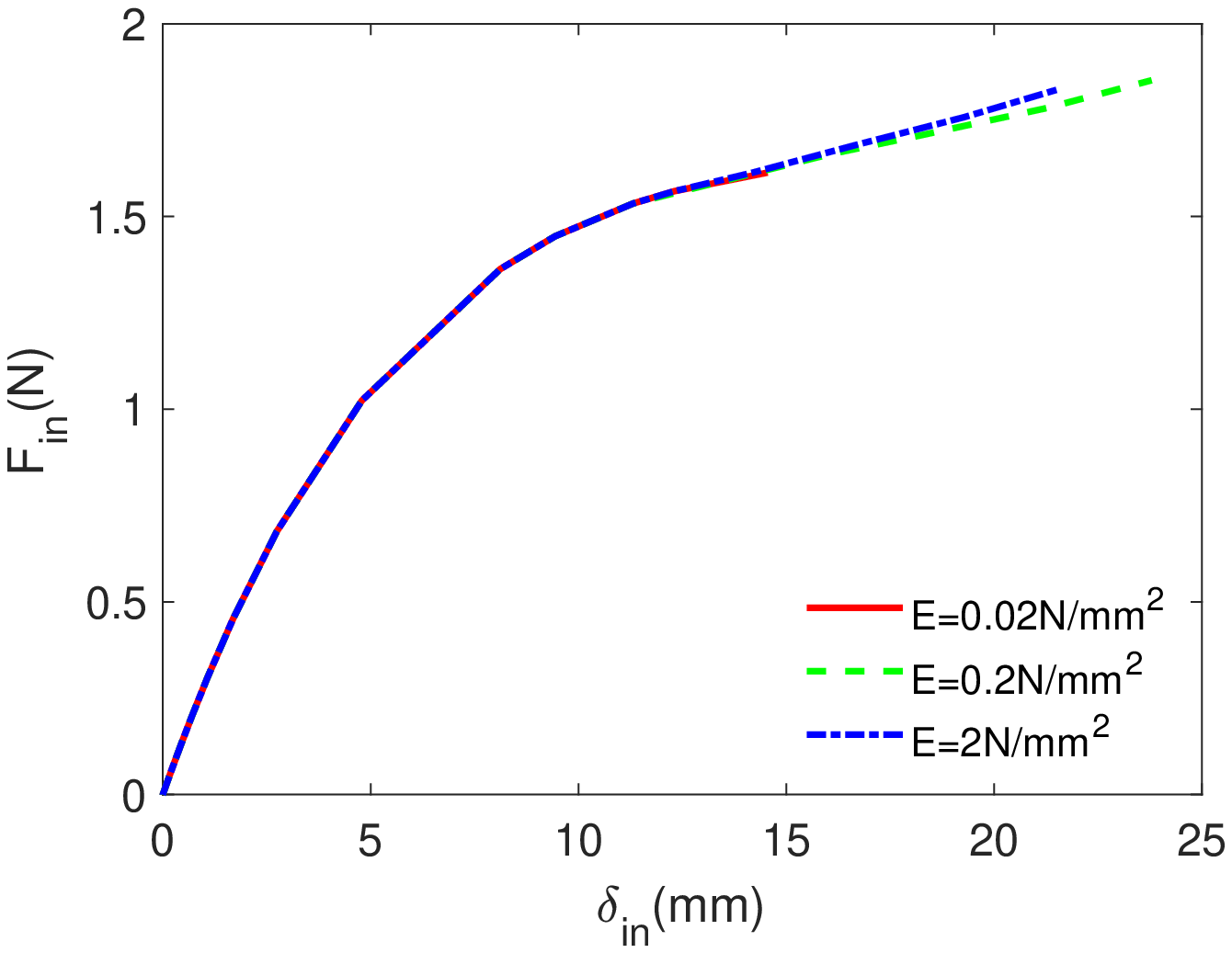}
		\caption{CiFM-I}
		\label{fig:CiFM1_wps_elliptical_various_Es}
	\end{subfigure} 
	\begin{subfigure}[t]{0.23 \textwidth}
		\centering
		\includegraphics[scale=0.3]{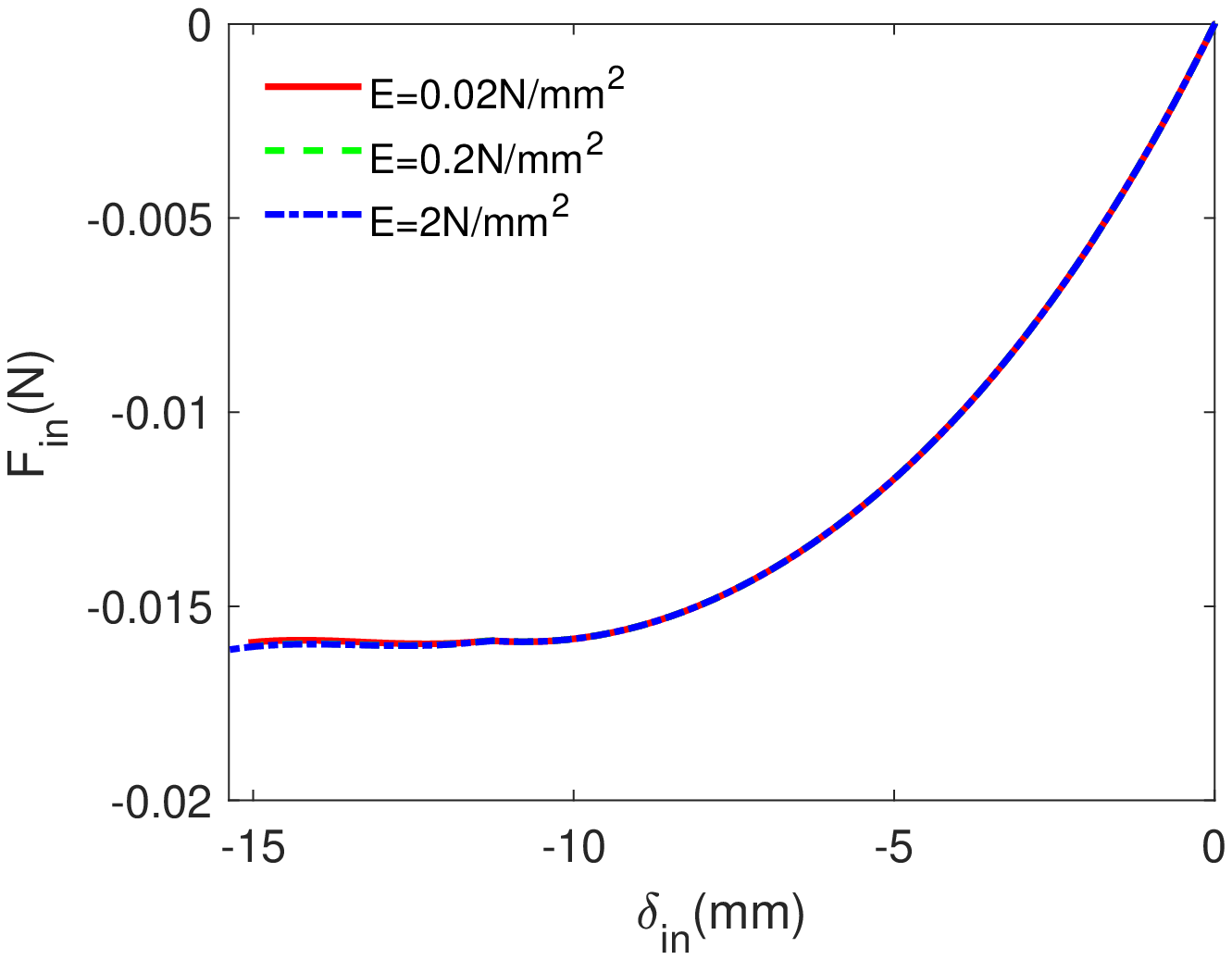}
		\caption{CiFM-II}
		\label{fig:CiFM2_wps_elliptical_various_Es}
	\end{subfigure} 
	\caption{Force-deflection characteristics for (\subref{fig:CoFM1_wps_elliptical_various_Es}) CoFM-I, (\subref{fig:CoFM2_wps_elliptical_various_Es}) CoFM-II, (\subref{fig:CiFM1_wps_elliptical_various_Es}) CiFM-I and   
		(\subref{fig:CiFM2_wps_elliptical_various_Es}) CiFM-II when the elastic modulus of the workpiece is changed. Workpiece is elliptical in shape, and major and minor axes dimensions are the same as the respective sides of the rectangular workpiece in Fig. \ref{fig:CFMs_f_d_curves_rectangular_wp_props}. }
	\label{fig:CFMs_f_d_curves_elliptical_wp_props}
\end{figure*}

In general, force-deflection characterstics of Constant output and Constant input force mechanisms can both change with the properties and shapes of the workpiece. Figure \ref{fig:CFMs_f_d_curves_rectangular_wp_props} depicts force-deflection response of the four continua when Young's modulus of the workpiece is changed. The workpiece is still rectangular in shape. For CoFM-I and CoFM-II, as the workpiece becomes stiffer, the output force required is large (Figs \ref{fig:CoFM1_wps_rectangular_various_Es} and \ref{fig:CoFM2_wps_rectangular_various_Es}). Further, CoFM-I tends to loose the constant output force characteristics. One observes that portion of contact region between CoFM-I and the continuum is lost --- a gap gets created between the two, and only regions near the left top and bottom corners of the workpiece are in contact. In case of CoFM-II (Fig. \ref{fig:CoFM2_wps_rectangular_various_Es}), the overall constant output force characteristic is maintained, and the magnitude gets slightly increased with workpiece stiffness. Nature of the input force in case of CiFM-I (Fig. \ref{fig:CiFM1_wps_rectangular_various_Es}) does get affected adversely. For CiFM-II (Fig. \ref{fig:CiFM2_wps_rectangular_various_Es}), however, nature and magnitude of the input force remain, desirably, unaltered. With elliptical (change in shape and properties of the) workpieces, CoFM-I and CoFM-II almost maintain constant output force characteristics (Figs. \ref{fig:CoFM1_wps_elliptical_various_Es} and \ref{fig:CoFM2_wps_elliptical_various_Es} respectively), but with changed force magnitudes. Constant input force characteristics for CiFM-I (Fig. \ref{fig:CiFM1_wps_elliptical_various_Es}) get thoroughly disturbed while those for CiFM-II (Fig. \ref{fig:CiFM2_wps_elliptical_various_Es}) are still maintained.

\subsection{Choice for the search algorithm}
\label{sec:srch_algo_choice}

Some of the previous methods \cite{Liu-et-al-2020,Liu_Chung_Ho_2021} employ gradient based search to synthesize CoFMs. They employ additive hyperelasticity	techniques by adding hyperelastic elements so that the finite element analysis is numerically stable. They further use filtering and projection so that solutions are free from checkerboards and that grey regions are minimized. Notwithstanding the time taken for synthesis, which is quite the case for all examples presented herein, we use stochastic Random Mutation Hill Climber search, detailed in \cite{Reddy-saxena-2020}, for multiple reasons. Primarily, we model \emph{contact} in our finite element analyses to cater to all possible scenarios during synthesis --- buckling of member(s), members interacting with themselves (self contact), and/or with external surfaces of specific shapes (mutual contact), and importantly, interaction of the continuum with workpiece to compute contact force (transfer). Computation of the interactive (output) force between the continuum and workpiece is a specific case of  mutual contact. Per se, these contact interactions are binary/discontinuous in nature (they either exist or not) and can introduce non-differentiability due to which sensitivity computations can become very difficult and/or cumbursome. Further, we aim to synthesize solutions that are manufacturable \emph{as is}, one reason why we decouple topology and shape/size optimization variables (they are coupled in gradient search), model the former as discrete and the latter as continuous variables. Limits on continuous design variables permit us to impose length scale constraints. Solutions that are non-convergent, those for which there is no force transfer, and other such are penalized. We avoid solutions with very thin members that by themselves can undergo significant deformation, and their presence or absence may cause alterations in the overall deformation characteristic. In essence, we avoid situations wherein optimized solutions need to be interpreted subjectively, e.g., removal of very thin regions, filleting around point connections, interpretation of gray cells, etc. We however remove dangling regions that we know do not store any strain energy during the entire deformation history.

\begin{figure*}[h!]
	\centering
	\begin{subfigure}[t]{0.24\textwidth}
		\includegraphics[width=\textwidth]{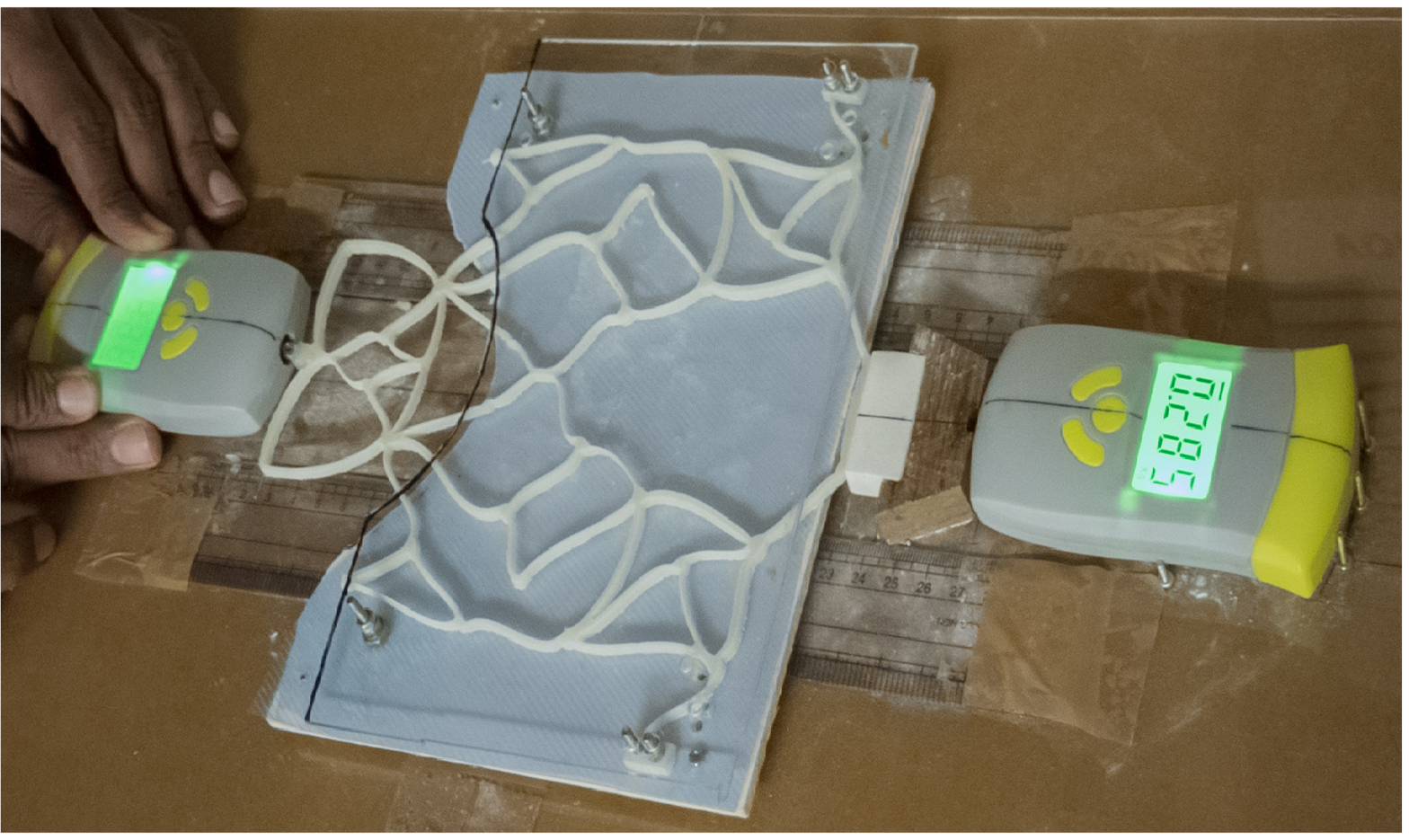}
		\caption{}
		\label{fig:CoFM1_exp_setup}	
		\includegraphics[width=\textwidth]{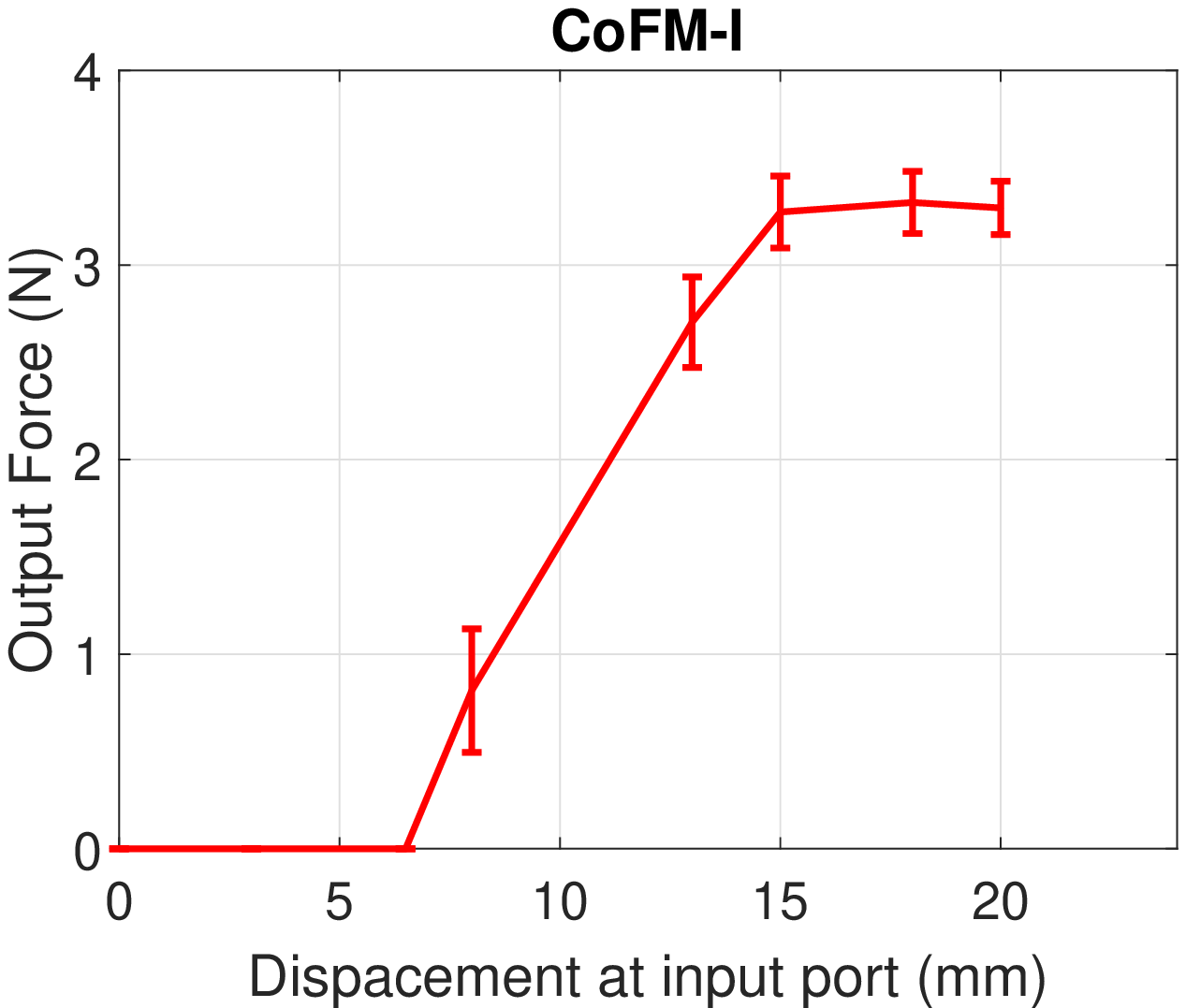}
		\caption{}
		\label{fig:CiFM1_exp_plot}
	\end{subfigure}	
	\begin{subfigure}[t]{0.24\textwidth}
		\centering
		\includegraphics[width=\textwidth]{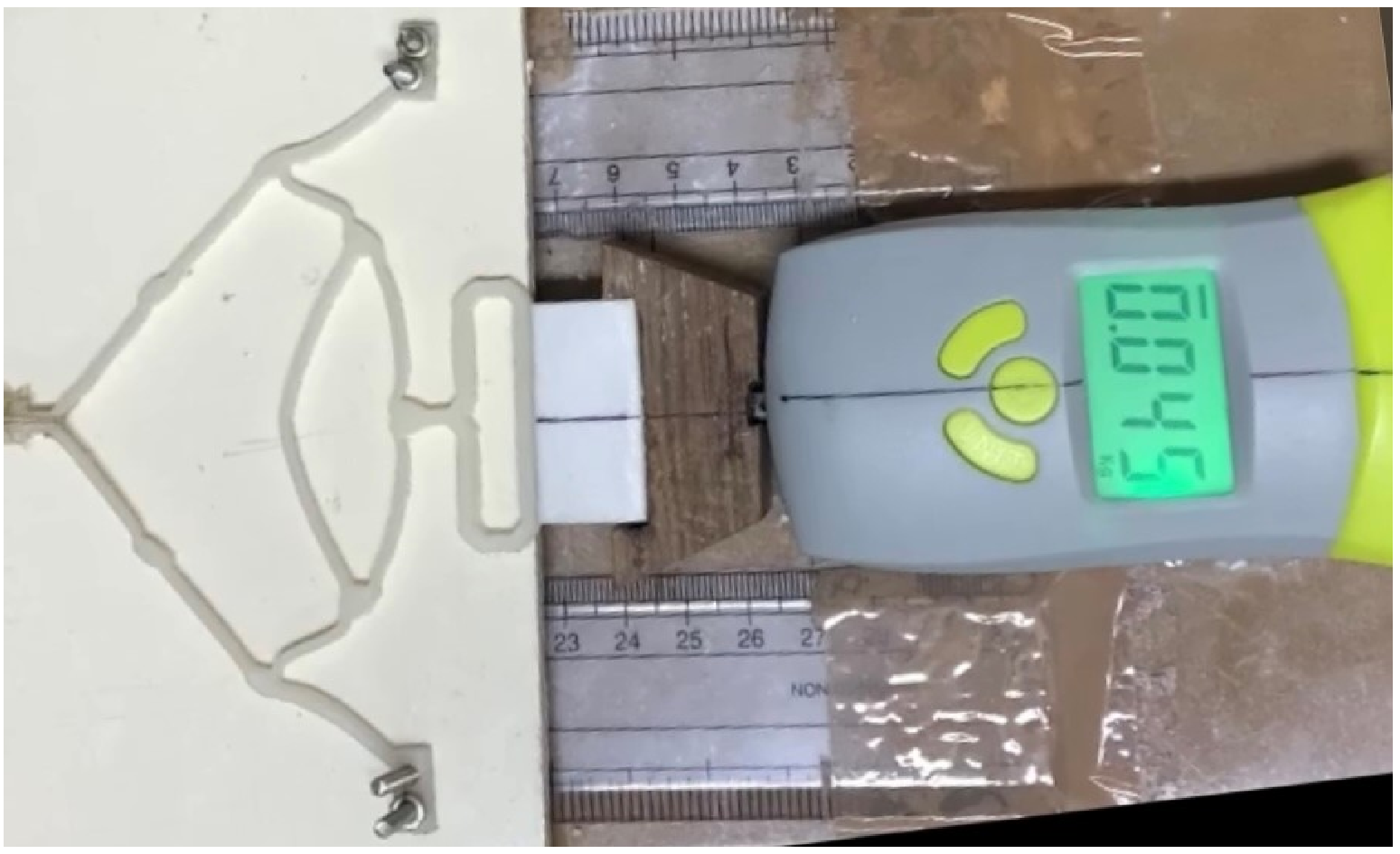}
		\caption{}
		\label{fig:CoFM2_exp_setup}
		
		\includegraphics[width=\textwidth]{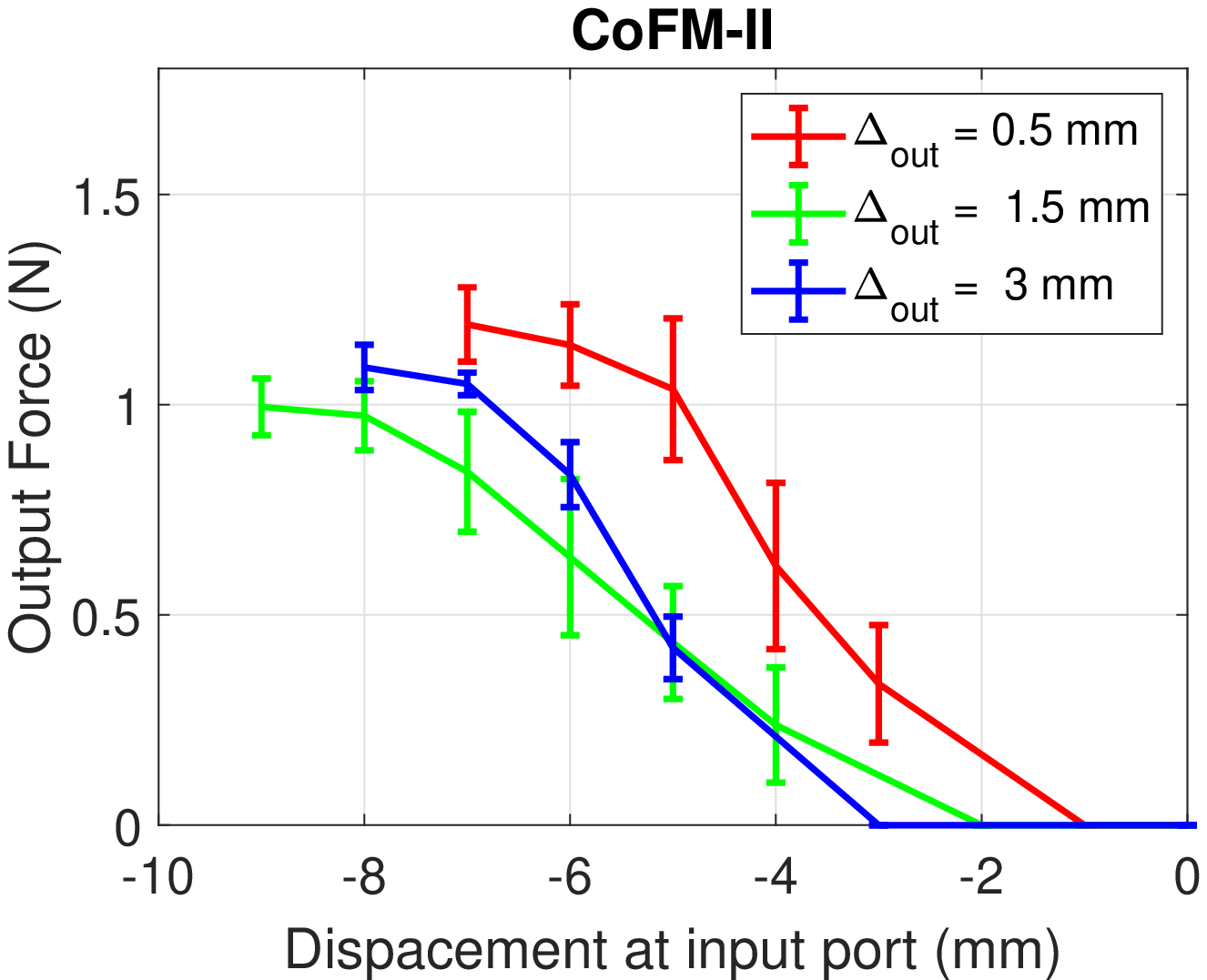}
		\caption{}
		\label{fig:CiFM2_exp_plot}
	\end{subfigure}	
	\begin{subfigure}[t]{0.24\textwidth}
		\centering
		\includegraphics[width=\textwidth]{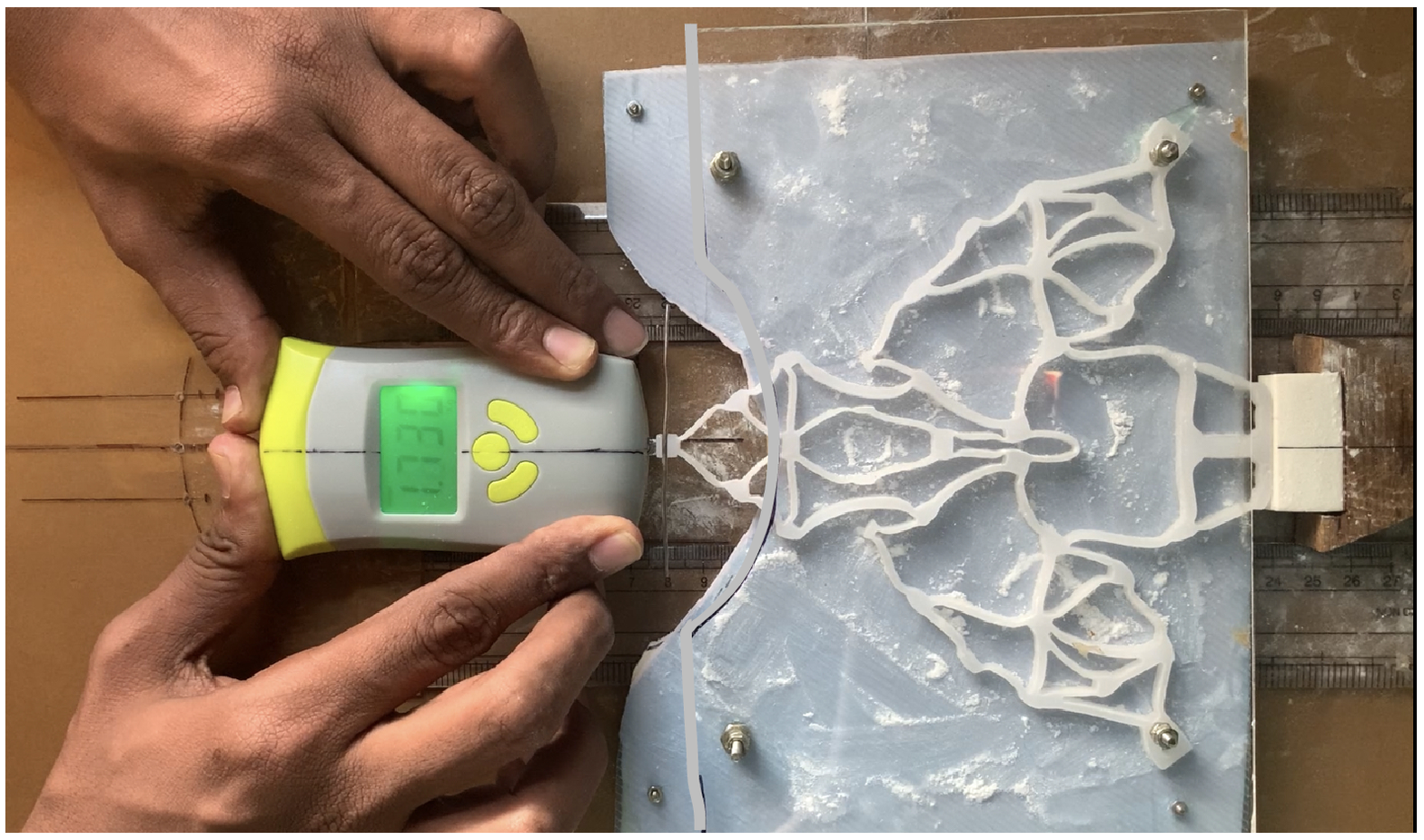}
		\caption{}
		\label{fig:CiFM1_exp_setup}	
		\includegraphics[width=\textwidth]{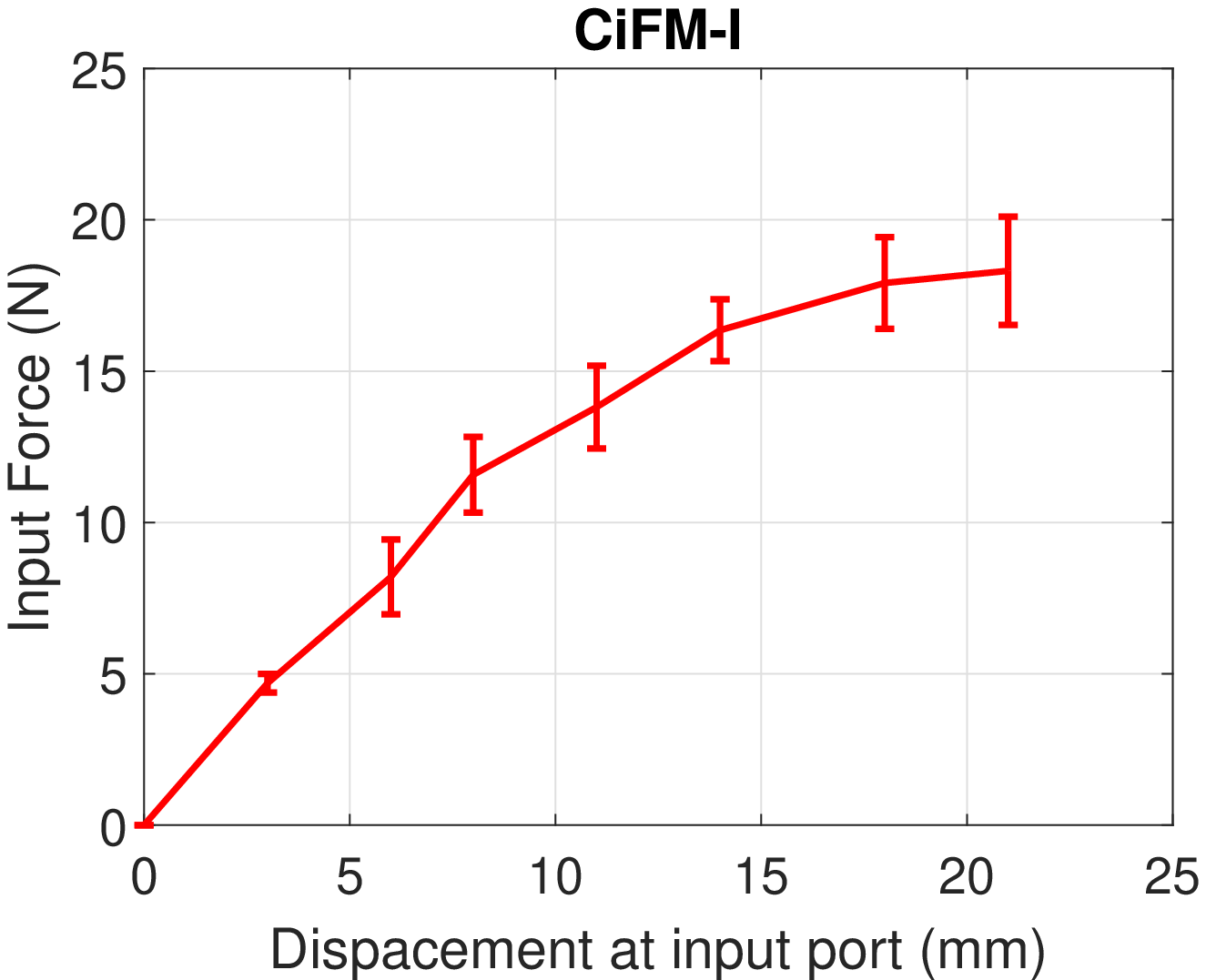}
		\caption{}
		\label{fig:CoFM1_exp_plot}
	\end{subfigure} 	
	\begin{subfigure}[t]{0.24\textwidth}
		\centering
		\includegraphics[width=\textwidth]{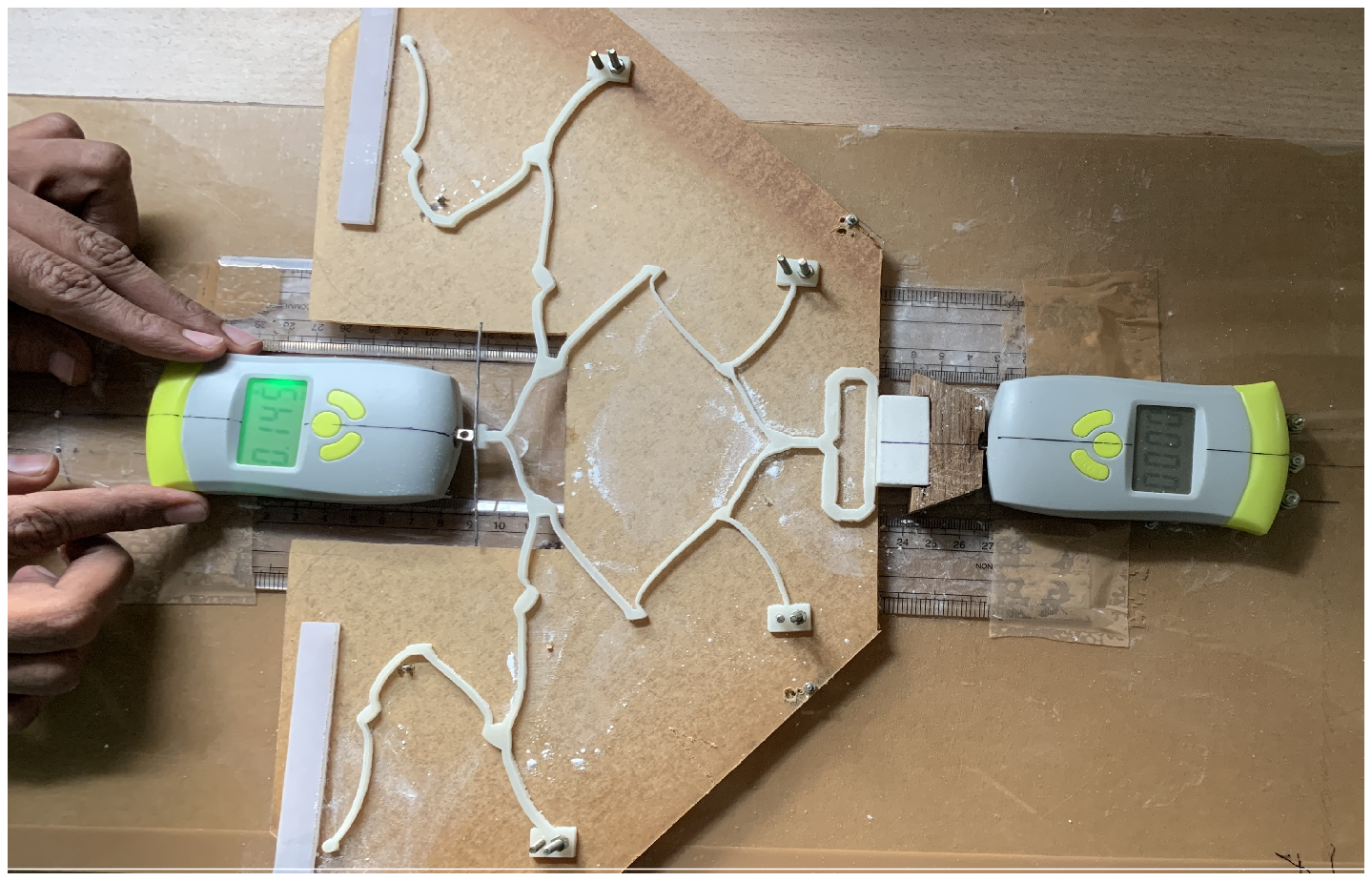}
		\caption{}
		\label{fig:CiFM2_exp_setup}	
		\includegraphics[width=\textwidth]{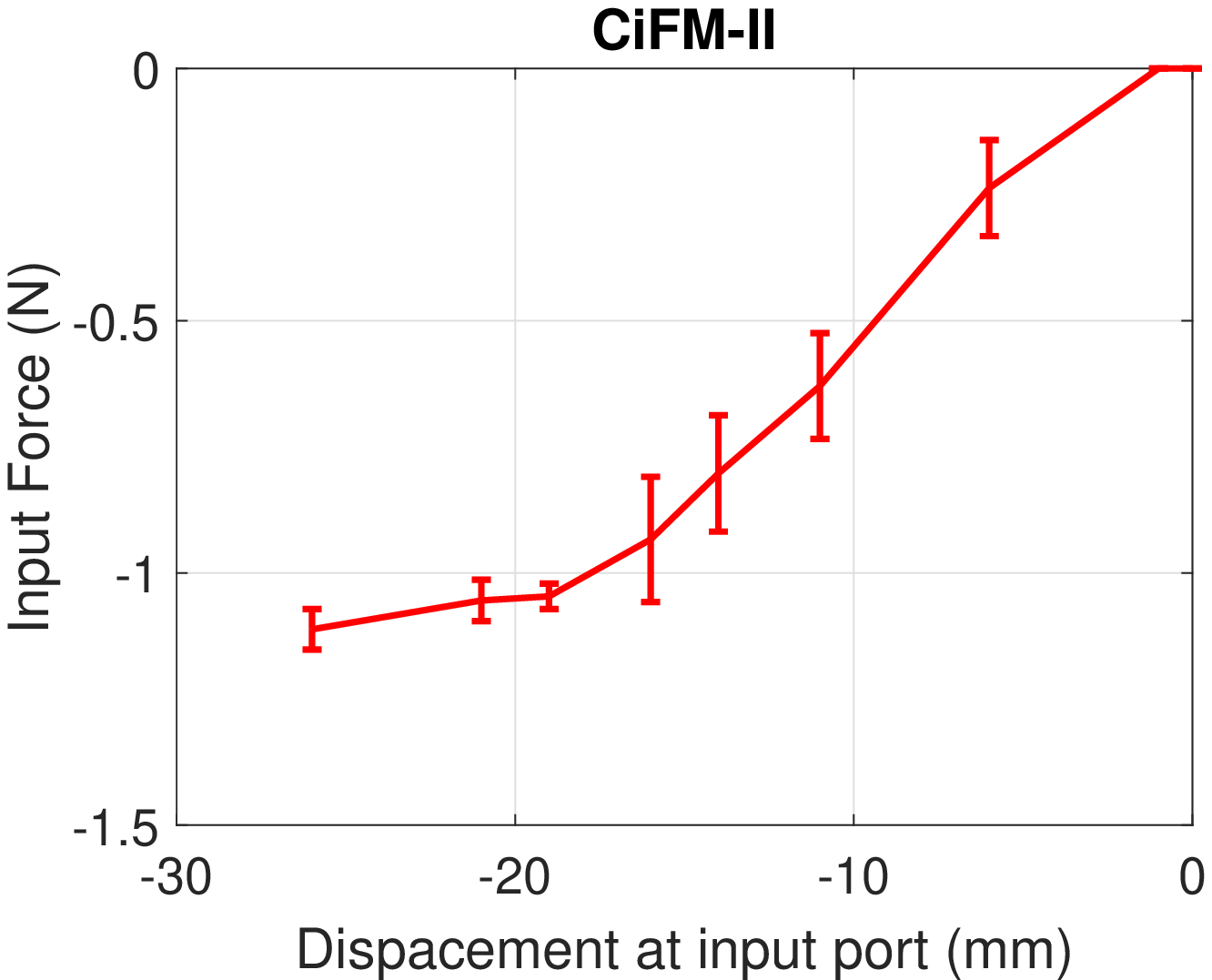}
		\caption{}
		\label{fig:CoFM2_exp_plot}
	\end{subfigure}
	
	\caption{(\textit{top row}) Experimental setup for CoFM-I/II and CiFM-I/II. Dimensions and distances between output port and the workpiece ($\Delta_{\textmd{out}}$) for each mechanism are taken from synthesis ($\Delta_{\textmd{out}}:$ 4.8mm, 4.5mm, 4.6mm and 1.9mm for CoFM-1 and II, CiFM-I and II respectively). Force is applied at the input port by manually pushing/pulling the input gauge with constant speed. Mechanisms are actuated multiple times to test repeatability, and mean values are used. A camera is mounted on the top to record the force-displacement data. (\textit{bottom row}) Experimental force-displacement response of the respective mechanisms. Curves use mean values and standard deviations (error bars). In CoFM-I, the output force is near constant for input displacement range of 15-20 mm. CoFM-II, for different gaps ($\Delta_{\textmd{out}}$), shows limited displacement before it gets locked, after which any further input force does not cause output displacement. However, all three curves in CoFM-II, show 'flatter' trends for the output force towards the end. CiFM-I shows near constant input force for displacement value crossing 18 mm. CiFM-II assumes nearly constant input force for displacement range of 19-26 mm.}
	\label{fig:experimental_setup_plots}	
\end{figure*}

\subsection{Fabrication and testing}
\label{sec:fab_and_test}
To validate the constant input/output force characteristics of the synthesized CoFMs and CiFMs, we fabricated and tested them. All four mechanisms were fabricated with Polypropylene sheets using water jet cutting. To facilitate fabrication of designed mechanisms, we chose to scale CoFM1 and CiFM2 by 1.4 times their actual sizes. The respective experimental setups are shown in Fig. \ref{fig:experimental_setup_plots} (top row). The setup includes a base plate, two calibrated push-pull type force gauges, one each for input and output port, a Vinyl eraser as workpiece, workpiece holder and two rulers parallel to the input port to measure the input displacement. Fine grained talcum powder is used to minimize  friction between the mechanisms and the base plate. CiFM-I and CoFM-I exhibit little out of plane motion during actuation to arrest which,  a transparent plate is mounted on top of these mechanisms, as shown in Fig. \ref{fig:CoFM1_exp_setup} and Fig. \ref{fig:CiFM1_exp_setup}. The output force measuring gauge is fixed at a specified distance using firm screws. For all mechanisms, the push/pull type force is applied manually and as gradually (quasistatically) as possible. Force and displacement data is recorded using top mounted cameras.  Experiments are repeated a number of times for each of the mechanisms.

The mean force-displacement plots for the mechanisms are shown in Fig. \ref{fig:experimental_setup_plots} (bottom row). The error bars represent standard deviation in repeated experiments. High errors could be attributed to plastic deformation due to repeated actuation, friction, and manual recording of displacement data. The mean force-displacement curves for all four mechanisms show near constant input/output forces over a range of input port displacement exhibiting good agreement in behavior with the synthesized mechanisms (in Sec. \ref{sec:to_eg_cofm} and Sec. \ref{sec:to_eg_cifm}). Difference in magnitudes of constant forces and displacement ranges could be attributed to different material properties of the mechanisms and workpieces used in synthesis and experiments. CoFM-I generates almost constant force for input displacement range of 15 mm to 20 mm. CoFM-II is able to deform till about 7mm to 9mm of input displacement. Thereafter, it locks in a configuration similar to that in Fig. \ref{fig:CoFM_Symmetric_example2_e}, and any further input displacement is barely possible. To investigate further, we tried three different gaps between the workpiece and the output port --- $\Delta_{\textmd{out}} = 0.5, 1.5, \text{ and } 3 $ mm. The trend remained similar with the output force almost attaining a constant value for the three cases. Applying very high input force after the locked position resulted in rupture of the mechanism. In CiFM-I, one observes a near constant input force of 18N for the input displacement range of 18-21 mm. CiFM-II assumes nearly constant input force for displacement range of 19-26 mm.

\label{sec:close_cofm_cifm}

A methodology to design constant input and output force mechanisms is presented. Contact modeling is employed not only to accurately model interactive forces with flexible workpieces, but also, to capture a variety of deformative and interative modes members of a compliant mechanism can undergo, thus generating many design possibilities. Novel objectives that decouple force magnitudes, near zero slope requirement and displacement range over which mechanisms exhibit constant force characteristics, are proposed and employed. Two examples each of constant input and output force mechanisms are presented. It is observed that presence of external contact surfaces (other than the workpiece) may not be mandatory to observe constant force characteristics. However, external surfaces can contribute to interesting solutions. Effect of change in shape and material properties of workpiece on force-displacement characteristics of CFMs is studied, and one finds that such characteristics do get altered.
Further, the synthesized CoFMs and CiFMs are validated via their respective prototypes. It is found that despite different material constitution, and associated plastic deformation, the mechanisms, by and large, possess the desired characteristics. Incorporating plastic deformation/ yield strength like criteria in the proposed design methodology is a possibility in future when considering the design of say, statically balanced compliant mechanisms.
		
\bibliography{References_new}

\bibliographystyle{ieeetr}

\end{document}